\documentclass[11pt]{article}

\pdfoutput=1
\usepackage {amssymb, amsmath}
\usepackage{graphicx}

\newcommand{\Om}{\mathcal{O}}
\newcommand{\sources}{\Sigma_n}
\newcommand{\sourcedest}{\mathcal{P}_n}
\newcommand{\system}{\mathcal{S}_n}
\newcommand{\hopset}{\mathcal{H}}
\newcommand{\DC}{\mbox{DC}}

\newcommand{\defeq}{\ensuremath{\stackrel{\triangle}{=}}}

\newtheorem{definition}{Definition}
\newtheorem{theorem}{Theorem}
\newtheorem{lemma}[theorem]{Lemma}
\newtheorem{corollary}[theorem]{Corollary}
\begin{document}
\setcounter{footnote}{1}

\begin{titlepage}
\begin{center}

\setlength{\baselineskip}{24pt}

{\LARGE   
On Throughput Scaling of Wireless Networks:\\ Effect of Node Density and
Propagation Model} 

\vspace{2em}

\setlength{\baselineskip}{18pt}
{\Large Enrique J. Duarte-Melo, 
Awlok Josan, Mingyan Liu,  David L. Neuhoff,\\ and Sandeep 
Pradhan}%

\vspace{2 em}

{\large Electrical Engineering and Computer Science Department\\
University of Michigan\\Ann Arbor, MI 48109}%

\vspace{2em} 

{\large Oct 18, 2006 }

\end{center}

\footnotetext[1]{\baselineskip=10pt 
This work was supported by NSF grants CCR-0329715 and
ANI-0238035.  Portions of this work were presented at
the IEEE International Symposium on Information Theory,
Seattle, July 2006.}

\setcounter{footnote}{1}

\vspace{2em}

\begin{abstract}
\setlength{\baselineskip}{18pt}
{\normalsize This paper derives a lower bound 
to the per-node throughput achievable
by a wireless network when $n$ source-destination pairs
are randomly distributed throughout a disk of radius $n^\gamma$,
$ \gamma \geq 0$, propagation is modeled by
attenuation of the form $1/(1+d)^\alpha$, $\alpha >2$, and successful 
transmission occurs at a fixed rate $W$ when received signal to noise
and interference ratio is greater than some threshold $\beta$, and at rate
0 otherwise.  The lower
bound has the form $n^{1-\gamma}$ when $\gamma < 1/2$, and
$(n \ln n)^{-1/2}$ when $\gamma \geq 1/2$.  The methods
are similar to, but somewhat simpler than, those in the seminal
paper by Gupta and Kumar.}
\end{abstract}

\end{titlepage}

\setlength{\baselineskip}{15pt}
 \section{Introduction} \label{sec-introduction}

The pioneering work of Gupta and Kumar \cite{gupta:march2000} has led to many
studies of scaling laws for the asymptotically achievable throughput in
wireless networks under a variety of network models and
assumptions. Such scaling laws help us understand the fundamental
performance limits of these networks and how
efficiency changes as network conditions change.
Some examples include
\cite{grossglauser:april2001} where the nodes are allowed to move;
\cite{duarte:2003,marco:april2003,heshamelgamal}, where
many-to-one type of communications is considered;
\cite{xie:kumar,heshamelgamal}, where cooperative communication
schemes are employed to improve network throughput; and
\cite{servetto_antenna}, where scaling laws are derived using
directional antenna assumptions. Other examples can
found in the June 2006 Special Issue on Networking
and Information Theory of the \emph{IEEE Trans.
Inform. Theory}.

All of these scaling results are highly dependent on the various 
assumptions made, such as on 
the network topology (e.g., planar, linear, ring, sphere, etc.), the 
purpose of the network (e.g., many-to-many vs. many-to-one 
communications), the physical layer models (e.g., different signal 
propagation and interference models), and the
asymptotic density of nodes (e.g. increasing to infinity or
remaining constant). 

This paper focuses on two such aspects --  
the underlying model for signal propagation and the
asymptotic density of nodes.   
We focus on the  many-to-many communications task.  Specifically, 
a set of $n$ nodes are randomly distributed over some region $A$, and
each node randomly chooses another node to whom to transmit data.  
All such transmissions use the same power $P$, which the designer can choose,
and communicate bits at some fixed rate $W$ that does not depend
on $P$.
Transmissions are received in the presence of interference from
other nodes transmitting at the same time, as well as from background noise.
They are modeled as successful if the signal
to interference and noise ratio (SINR) at the receiver is above some threshold
and unsuccessful if not.
For this task, \cite{gupta:march2000} found the maximum attainable
throughput per node
is\footnote{We use the notation $\Om(f_n)$, $\Omega(f_n)$ and
 $\Theta(f_n)$, in the conventional way, i.e., to characterize a 
 quantity $x_n$ depending on $n$ for which
 there are finite constants  $c>0$, $d>0$ and $n_0$ 
 such that, for all $n>n_0$, respectively, $x_n < d f_n $, $x_n > c
 f_n$, and $c f_n < x_n < d f_n$.} 
$\Omega \big( {1 \over \sqrt{n \ln n} } \big)$  and $\Om \big( {1 \over \sqrt{n} } \big)$
bits/sec,
assuming a propagation law in which received power decays as $1 \over d^\alpha$ 
with transmission distance $d$, for some $\alpha > 2$.  That is, it scales
at least as $1 \over \sqrt{n \ln n}$, but no larger than $1 \over \sqrt{n}$. 

While \cite{gupta:march2000}
assumed that the $n$ nodes were randomly distributed over a fixed region $A$,
and consequently, the network becomes denser as $n$ increases, 
the maximum throughput is actually 
independent of the size of the region.  For example, it does not change if the
region size scales with $n$, as we will wish to consider in this paper.  
To see this, consider a specific set
of nodes transmitting simultaneously in some region $A$, each with 
power $P$, and each to its own receiving node.  
Now suppose the positions of all transmitting and receiving nodes are 
scaled by a factor $\mu$, and the transmit power is scaled by the 
factor  $\mu^\alpha$.  Then the SINR
(to be defined in Section \ref{sec-propagation}  in the obvious way) will be the same
at each of the scaled receiving nodes, as it was at each of the original unscaled
nodes.  It follows that the maximum attainable throughput is not
affected by a scaling of the region over which the nodes are distributed.
For example,  the $\Omega \big( {1 \over \sqrt{n \ln n} } \big)$ throughput law 
applies equally when the region $A$ is fixed and the density
of nodes increases linearly with $n$, or when the density of nodes
is fixed and area of $A$  increases linearly with $n$.

In contrast, Arpacioglu and Haas \cite{haas} 
have shown that when the
propagation model has the form $1 \over (1+d)^\alpha$, for some $\alpha > 2$, and 
the region $A$ remains fixed (so node density
increases linearly with $n$), the maximum attainable throughput
decreases dramatically to  $\Theta \big( {1\over n } \big)$, 
which is the throughput attained by simple time sharing among the $n$ nodes.  
On the other hand, the two propagation models are essentially equivalent in the far field. 
As a result, if the area of $A$ increases at least linearly with $n$,  then because
the distances between nearest nodes are not decreasing with $n$ to zero,
it is relatively easy to see that  the maximum throughput is the same for both 
models, i.e., it is $\Omega \big( { 1 \over \sqrt{n \ln n} } \big)$ and 
 $\Om \big( {1 \over \sqrt{n} } \big)$.
The above cited results on the maximum attainable throughputs are summarized in 
Table \ref{T:throughputs}. 
One concludes that throughput depends significantly 
on the assumptions about  propagation model and node density.

\vspace{1 em}

\begin{table}[h]
\begin{center}
\begin{tabular}{c||c|c}
 & \multicolumn{2}{c}{Propagation Models}     \\
  & $1 \over d^\alpha$   & $1 \over (1+d)^\alpha$   \rule[-.8em]{0cm}{2em} \\
\hline \hline
fixed area ($\gamma=0$)
    & $\Omega \big( \frac{1}{\sqrt{n \ln n}} \big)$, $\Om \big( {1 \over \sqrt{n} } \big) $  
       \cite{gupta:march2000}   \rule[-.8em]{0cm}{2em}
    & $\Theta \big( \frac{1}{n} \big)$ \cite{haas} \\
\hline
fixed density ($\gamma={1 \over 2}$)
  &$\Omega \big( \frac{1}{\sqrt{n \ln n}} \big)$, $\Om \big( {1 \over \sqrt{n} } \big) $
     ~~~~
  &$\Omega \big( \frac{1}{\sqrt{n \ln n}} \big)$, $\Om \big( {1 \over \sqrt{n} } \big) $
     \rule[-.8em]{0cm}{2em} \\
\hline
\end{tabular}
\caption{Throughput scaling results for random networks under 
different propagation models and 
network density assumptions.}\label{tbl-summary} \label{T:throughputs}
\end{center}
\end{table}

\vspace{-1em}

In this paper, we focus on the $1 \over (1+d)^\alpha$ propagation model and
the gap, evident in the rightmost column of Table \ref{T:throughputs},
between the maximum throughputs attainable for fixed area
and fixed density.  Specifically, we ask how attainable throughput
changes as the node deployment scenario ranges from fixed area
to fixed or decreasing density.  
We do this by considering the network region
$A$ to be a disk with radius  $n^\gamma$, where $\gamma \geq 0$ is 
a parameter that determines the deployment scenario.  
The choice $\gamma = 0$ corresponds to a network with fixed area and node
density increasing linearly with $n$.  The choice $\gamma = {1 \over 2}$ corresponds
to a network with area increasing linearly with $n$ and density remaining
constant.  Intermediate values of $\gamma$ correspond to the network
density increasing sublinearly, while $\gamma>{1 \over 2}$ corresponds
to decreasing network density.
We consider time-slotted systems 
and measure throughput in bits/slot, which of course can be easily converted
to bit/sec.  The principal result of the paper is that throughput 
$\Omega \big( {1 \over n^{1-\gamma} } \big) $  is attainable when
$\gamma < {1 \over 2}$, whereas throughput 
$\Omega \big( { 1 \over \sqrt{n \ln n} } \big) $ is attainable 
when $\gamma \geq {1 \over 2}$.  
If it is desired to measure throughput in bit-meters/slot, then these
results are multiplied by $n^\gamma$.

For $\gamma = 0$,
the attainable throughput  $\Omega \left(1 \over n^{1-\gamma} \right) $  
is consistent with the 
$\Theta \big({1 \over n } \big)$ result  found in \cite{haas}.  
As $\gamma$ increases towards $1 \over 2$, the attainable throughput 
$\Omega \big( {n^\gamma \over n} \big)$
increases, due essentially to the fact that as $\gamma$ increases, 
there is room for more simultaneous transmitters.
For $\gamma \geq {1 \over 2}$, the attainable throughput scaling rate saturates at 
$\Omega \big( { 1 \over \sqrt{n \ln n} } \big)$.
This is the rate  found for $\gamma = 0$ and
the $1 \over d^\alpha$ propagation model \cite{gupta:march2000}, 
that also applies to $\gamma = {1 \over 2}$ 
and the $1 \over (1+d)^\alpha$ propagation model (see Table \ref{T:throughputs}).
For $\gamma = {1 \over 2} - \epsilon $ and very small $\epsilon$,
one might be tempted to interpret the result as saying
that throughput 
$\Omega \big( {1 \over n^{1-\gamma}} \big)   
\approx \Omega \big( { 1 \over \sqrt{n}} \big)$  is attainable,
which would be larger than the attainable throughput 
$\Omega \big( { 1 \over \sqrt{n \ln n} } \big)$ for $\gamma = {1 \over 2}$,
and would contradict the notion that attainable throughput 
does not decrease when $\gamma$ increases.  However, the result
actually says the attainable throughput is $\Omega \big( {1 \over \sqrt{n} \, n^\epsilon} 
\big)$, which is a smaller lower bound than 
$\Omega \big( { 1 \over \sqrt{n} \, \sqrt{ \ln n} } \big)$,
the attainable throughput for $\gamma = {1 \over 2}$, no matter how 
small $\epsilon$ is.

Interestingly, Franceschetti et al. \cite{franceschetti:d:t:t:04} have shown
recently that larger throughput,
$\Omega \big( { 1 \over \sqrt{n} } \big)$, is attainable in  a variety
of situations.  These include the $1 \over d^\alpha$ propagation
model and both a fixed area ($\gamma=0$) and a fixed density 
($\gamma = {1 \over 2}$) network region.  They also include a fixed
density network and a propagation model that is bounded, like 
the $1 \over (1+ d)^\alpha$ propagation model considered in the present paper. 
For the $1 \over d^\alpha$ propagation model, the previously mentioned
 invariance of SINR to dimension scaling of the network region
and appropriate scaling of power implies that throughput 
$\Omega \big( { 1 \over \sqrt{n} } \big)$
is in fact attainable  for all $\gamma \geq 0$.   For the bounded
propagation model, it is not evident what happens when $\gamma < {1 \over 2}$.
The larger throughputs demonstrated in \cite{franceschetti:d:t:t:04} are 
obtained assuming that the rate of successful transmission
between two nodes equals the capacity of an additive Gaussian channel 
with signal to noise ratio equal to the received SINR.  This contrasts with 
the two-rate transmission assumed in \cite{gupta:march2000,haas}
and the present paper, 
in which the rate is $W$ when received SINR exceeds a threshold
and 0 otherwise. 
The construction in \cite{franceschetti:d:t:t:04}  also adopted a hierarchical 
structure where
packets are first sent to a backbone from which they are routed to the
destination.  This contrasts with the straight line shortest path type of routing
used in \cite{gupta:march2000,haas} and the present paper.

Assuming the $1 \over (1+d)^\alpha$ propagation law,  $\gamma \geq 0$,
and the two-rate transmission model, 
it may well be that throughput cannot scale at rates above those
we show to be attainable.  However,
no such proof or claim is offered in this paper.

In the remainder of the paper, 
Section \ref{sec-manytomany} introduces the many-to-many
communication task, along with 
a concrete specification of a system for this task, its throughput
and the notion of a successful system.  The latter is determined
by a propagation model and a criterion for judging the success of 
a transmission in the presence of interfering transmitters and background 
noise. 
The specific success criterion and propagation model used in this paper are
introduced in Section \ref{sec-propagation}.
Section \ref{sec-success} introduces distance-based success criteria,
which are like the protocol models used in \cite{gupta:march2000}, and it 
discusses their relationship to the SINR-based 
physical model of \cite{gupta:march2000}.
Section \ref{sec-positive} states and proves the main result.   
Section \ref{sec-conclusions} summarizes and makes concluding remarks. 
Finally, a few details are relegated to appendices.  

While the methods used here are related to those used in previous work 
(e.g., the use of distance-based protocol models for determining when 
a set of simultaneous transmitters will not interfere with each other 
\cite{gupta:march2000}, and the use of straight lines intersecting
cells of a partition to determine routes),
they differ in key respects (e.g.,  the
dividing of the load as equally as possible among the nodes
within a partition cell, and the use of the Chernoff bound instead
of uniform convergence of the weak law of large numbers).
As one benefit, the new methods permit straightforward analysis of
throughput scaling on a disk, rather than the surface of a sphere,
despite the hot-spot-at-the-center problem. 
In addition, we clarify the role of protocol models, and their 
relations to physical models, in aiding the design of a system
and the demonstration of attainable throughputs. To illustrate the
generality of our methods, we also indicate in Section \ref{sec-positive}
how they can straightforwardly 
demonstrate the original $\Omega \big( {1 \over \sqrt{n \ln n} } \big)$ throughput
result of \cite{gupta:march2000}.

\section{The Many-to-Many Communication Task} \label{sec-manytomany}

\vspace{-.25em}
A set of $n$ nodes,  $\sources = \{s_1,\ldots,s_n\}$, is distributed
over a  disk $A_n \subset \mathcal{R}^2 $  with  radius  $n^\gamma$,
called the \emph{network region},    
where  $s_i \in A_n$  denotes the 
location within the disk of the $i$th node,  where   
$\gamma  \geq 0$  is a fixed  
parameter that characterizes how the area of the disk and the density of nodes
scale with $n$.   Each node serves as
a source of bits that it wishes to communicate to some destination. 
For each source  $s_i$,  another of the $n$  nodes, denoted  $d_i$,  
is designated as the \emph{destination} for its bits.  As a result, there is a 
\emph{source-destination set} $\sourcedest =  \{ (s_1,d_1), \ldots, 
(s_n,d_n) \} $  
consisting of $n$  source-destination pairs, each representing a desired
\emph{conversation}.  Note that a node may serve 
as the destination for more than one source.

Each of the $n$ sources has an infinite number of bits it wishes to communicate 
to its destination node, as quickly as possible.  Communication uses simple 
multihop relaying with a time slotted system.  
We make the usual assumption that the source-destination set is random.  
Specifically,  $s_1, \ldots, s_n$   are drawn independently, each with a uniform 
distribution on the disk.  
Then for each $s_i$, the destination $d_i$ is equally likely to be any
$s_j, j \neq i$,  independent of all other $s$'s and $d$'s.  
(Two sources may have the same destination.)

Roughly speaking, for a given  $n$,  one wishes to find the largest number  
$\lambda$ such that for all source-destination sets $\sourcedest$,  
except a set with small probability,  $\lambda$  bits/slot can be successfully 
transmitted from each source to its destination.   In this paper we 
do not claim to have found the largest possible $\lambda$.
However, we are able to show that with 
$\lambda_n = \Omega \big( {1 \over n^{1-\gamma}} \big) $, $\gamma < {1 \over 2}$
and $\lambda_n = \Omega \big( {1 \over \sqrt{n \ln n} } \big) $, 
$\gamma \geq {1 \over 2}$, 
with probability approaching one as $n \to \infty$, there exist systems that send
$\lambda_n$  bits/slot.

\subsection{System Definition}

We now describe the kind of system to be used for the many-to-many task.
This is basically an explicit formalization of the kind of system that appears
in prior work. 
There is a transmitter and receiver at each of the $n$ nodes.  
The antennas at each node are omnidirectional. 
All transmitters use the same power $P$, which we get to choose and 
which may depend on  $n$  and the specific source-destination set  
$\sourcedest$.   (However, we will see in Theorem \ref{thm:main}
that when our system is optimized and $\gamma < {1 \over 2}$, 
the power $P$ can remain constant.) 
As mentioned earlier, transmissions occur in slots.  
We assume there is a fixed  $W > 0$  such that each transmitter can transmit at most  
one packet, consisting of $W$  bits, in one slot, regardless of  $P$, $n$ or any 
other factors.  Such transmissions are received throughout the network region
 $A_n$ in the 
presence of background noise with power $N_o$ and interference from other 
transmitters transmitting at the same time.
As a result, the packets might or might not be successfully received by an
intended receiver.  Criteria for determining success will be introduced later.
 
Each receiver can store an arbitrary number of packets, for later retransmission. 
However,  we will see there is  little need for such storage.
 
To communicate bits from the sources to their destinations, each
source-destination pair needs a \emph{route} and a \emph{schedule}.
A \emph{route} for source-destination pair $(s_i,d_i)$ is a finite sequence of
\emph{hops},  $h_i = (h_{i,1}, \ldots, h_{i,J_i})$, from $s_i$ to $d_i$  
with the $j$th hop of the route being a pair $h_{i,j} = (t_{i,j},r_{i,j})$ 
indicating that node $t_{i,j} \in \sources$ is to transmit bits originating 
at $s_i$ with the intention that they be received%
\footnote{We say ``intention'' because the omnidirectionality of the antennas
means that other nodes will also hear the transmission, and because noise or interference
may prevent the transmission from being successfully received.}
by node $r_{i,j} \in \sources $. 
The first hop 
has the form  $h_{i,1} = (s_i,r_{i,1})$,  subsequent hops have $r_{i,j} = t_{i,j+1}$, 
and the last hop has the form  $h_{i,J_i} = (t_{i,J_i},d_i)$.   
Since nodes are presumed to be able to indefinitely store packets 
received in previous time slots,  there is no need to allow routes 
to have loops, i.e.~for one node to appear twice in a route.  Accordingly we disallow 
loops, which implies that all  routes have length $n-1$ or less. 
Paths for different source-destination pairs may have different numbers of hops.
The \emph{length} of a hop $h = (t,r)$ is the Euclidean distance $\|t-r\|$.

A \emph{schedule} for route $h_i = (h_{i,1}, \ldots, h_{i,J_i})$ is a sequence  
of positive integers  $\sigma_i = (\sigma_{i,1}, \ldots, \sigma_{i,J_i})$  assigning 
a time slot to each hop of the route.  
Specifically, node $t_{i,j}$ makes its transmission of hop $h_{i,j}$ 
in time slot $\sigma_{i,j}$.  Combining the notions 
of route and schedule, each source-destination pair $(s_i,d_i)$ is assigned a 
\emph{scheduled route} 
 $H_i = ( (h_{i,1},\sigma_{i,1}), \ldots, (h_{i,J_i}, \sigma_{i,J_i}) )$.
 The hops in a route need not be assigned slots in increasing order; i.e.,
 we permit $\sigma_{i,j} > \sigma_{i,j+1}$.  

We now define a  \emph{system}  $\system$ for source-destination set $\sourcedest$
to be a set of $n$ scheduled routes 
$\{ H_1, \ldots, H_n \} $. 
Such a system is assumed to operate periodically
with period $p = \max_{i,j} \sigma_{i,j}$, which is the largest slot assignment of any
hop of any route.  That is, in steady state,  
the $j$th hop of route  $h_i = (h_{i,1}, \ldots, h_{i,J_i})$ 
is transmitted in slot $\sigma_j$ of each epoch of $p$ slots.   
The reason for restarting each route synchronously at the beginning of each
epoch will be explained shortly.

We also require scheduled routes of a system to be \emph{compatible} in the sense that 
no two hops, either from the same or different routes, can be scheduled to require 
transmission from the same
node in the same slot.   This requirement stems from our assumption that
a node can transmit at most once within a slot.  The previously stated assumption that 
all routes are transmitted again in every epoch of length $p$ (instead of, say, each
route cycling asynchronously) is designed to permit compatibility 
to be checked straightforwardly.  

In summary, a system $\system  = \{ H_1, \ldots, H_n \} $ for a set of 
source-destination pairs  $\sourcedest$
consists of  a compatible set of $n$ scheduled routes, one for  
each source-destination pair in $\sourcedest$,  
and with the latest time slot assigned to any hop being defined as 
the \emph{period} $p$  of  the system. 
For future use, for  $j \in \{1, \ldots, p \}$,  let us define 
the \emph{hop set}  $\hopset_j$ to be the set of hops $(t,r)$ that the system specifies 
as transmitting in the $j$th slot.
That is, $\hopset_j$ contains a hop  $h =(t,r)$  if  $h$ is a hop in some
scheduled route that is scheduled for the $j$th time slot. 
Let us also define the \emph{transmission set}  $T_j$ to be the
set of nodes that the system specifies as scheduled for the $j$th slot.
Note that for any $\sourcedest$, there obviously exists a set of routes,
and for any set of routes, one can always find a set of compatible
schedules for these routes.  For example, although not
very efficient, one could define a schedule
in which each hop of each route is assigned to a distinct slot.
Therefore, there always exists a system for any $\sourcedest$.

We now describe concretely how a system $\system = \{ H_1, \ldots, H_n \}$ with
period $p$ for source-destination set $\sourcedest = \{ (s_1,d_1), \ldots, (s_n,d_n) \}$ 
 transmits data from the sources to the destinations.  
For each $i \in \{1, \ldots, n \}$, the first packet from $s_i$ is transmitted via hop $h_{i,1}$ in
slot $\sigma_{i,1}$  of the first epoch of $p$ slots.  Then, if $\sigma_{i,2} > \sigma_{i,1}$, this packet is relayed via 
the second hop $h_{i,2}$, also in the first epoch.  
If not, it is transmitted in the second epoch.  Subsequently, for each $j>2$, the packet is
relayed via hop $h_{i,j} = (t_{i,j},r_{i,j})$ in slot $\sigma_{i,j}$ of the first epoch 
in which the packet has been received at $t_{i,j}$ prior to slot $\sigma_{i,j}$.  Moreover, 
transmission of subsequent packets from $s_i$ to $d_i$ are pipelined so that 
in steady-state, 
within each epoch, one new packet is generated by $s_i$, 
each hop $h_{i,j}$ of the corresponding scheduled route is executed once,
and one packet from $s_i$ (typically generated in an earlier epoch) 
is received at destination $d_i$.  And this happens
for each source-destination pair.  
Compatibility ensures that no node is asked to make two transmissions in one slot.

Notice that, as assumed earlier, there is no need for the  $\sigma_{i,j}$'s 
of a scheduled route
to be in increasing order.  A node simply stores each received packet until it is time to
transmit it, either in the present or next epoch.  As a result, it will never store more
than one packet at a time from one source-destination pair.

\subsection{Success Criteria}

We assume the existence of a \emph{transmission success criterion} 
that determines whether or not a given transmission will be successful.  
Specifically, when a node at 
location\footnote{Whereas
$s_i$ denotes the location of the $i$th node, letters $t$ and $r$
with one or no subscripts, such as $t_i$ and  $r_i$, are used as variables to 
denote the location of some transmitter and receiver, respectively.} 
$t$ transmits to a node at location 
$r$ in the presence of background noise with power $N_o$
and simultaneous transmissions from locations in the set
$T=\{t_1,t_2, \ldots,t_{M-1}\}$, 
the criterion determines  whether this communication is successful.
Such a criterion can be characterized by a \emph{success indicator function}
$\psi(t,r,T,P,N_o)$  of the form 
$\psi: \mathbb{R}^2 \times \mathbb{R}^2 \times \overline{\mathbb{R}^2} \times \mathbb{R}_+ \times \mathbb{R}_+ \rightarrow \{0,1\}$,
where $ \overline{\mathbb{R}^2}$ denotes the set of all finite subsets 
of $\mathbb{R}^2$,    $\mathbb{R}_+$  indicates the set of nonnegative
real numbers, and   $\psi(t,r,T,P,N_o) = 1$ indicates that conditions
are suitable for a successful transmission from $t$ to $r$ in the presence
of background noise with power  $N_o$ and simultaneous transmissions
from the locations in $T$, whereas $\psi(t,r,T,P,N_o)=0$
indicates they are not.

Although a success criterion in this general form can be used to model
communication in a variety of networks, since we are dealing with
wireless networks, we will use a success indicator criterion that is
characterized by a propagation model $\eta$ and a
power indicator function $\phi$. 
A  \emph{propagation model} is a
function $\eta: [0,\infty)  \rightarrow [0,\infty) $ such that  $\eta(d)$ 
determines the fraction of transmit power that is received at distance $d$ 
from the transmitter.  
A \emph{power indicator
function} is a binary function  $\phi(P',\{P_{1},P_2, \ldots,P_{M-1}\},N_o)$
that equals one if  the power
$P'$ received at $r$ from the  transmitter at $t$, the set of powers 
$\{P_1,P_2,\ldots,P_{M-1}\}$ received at $r$ from the other transmitters
at locations in $T$, and the background noise level $N_o$ are 
such that the transmission from $t$ to $r$ is successful, and zero if not.
We now restrict attention to success indicator functions of the form
\[
     \psi(t,r,T,P,N_o) = \phi ( P \eta(\|t-r\|), \{P \eta(\|t_1-r\|),
     \ldots, P\eta(\|t_{M-1}-r\|)\}, N_o ) ~. 
\]  

With a transmission success criterion in hand, one may now define a hop 
set $\hopset$ to be successful if for every hop $(t,r) \in \hopset$, transmission from
$t$ to $r$ is successful in the presence of transmissions from all other
transmitters in the transmission set $T$ corresponding to $\hopset$.  Next,
one may define a system to be successful if all of its hop sets are successful.
Note that in the situation described above in which a propagation model 
$\eta$ is available,  the success of a hop set or a system will depend on 
$P$, $N_o$ and the propagation model $\eta$, as well as the locations
of the transmitters and receivers of the hops in the hop set.

\subsection{Throughput}

If a system $\system$ with period $p$ for source-destination set $\sourcedest$  
is successful, i.e., if all hop transmissions are 
received successfully, then in steady-state the
system delivers one packet, consisting of $W$ bits, from each source $s_i$ to
its destination $d_i$ in each epoch of $p$ slots. Accordingly, we define
the  \emph{throughput} for a 
system 
to be 
\[
	\lambda = {W \over p} ~.
\]
Clearly, throughput is a meaningful quantity only when the system is successful.
To attain large throughput, one needs to design a set of compatible
scheduled routes with $p$ as small as possible.

It would have been possible to permit more than one route for each
source-destination pair.  In this case, one would define throughput to be 
$mW/p$, where $m$ is the minimum number of routes per
source-destination pair.  However, since this paper focuses only on finding 
a lower bound to achievable throughput, there is no need to consider such.
 
Note that the order in which hops in a route are scheduled has no effect on
throughput, though it will effect the delay until the first packets from each 
source appear at their destinations.

The problem of interest is to learn the order of the maximum possible throughput 
 that is attainable with high probability when $n$ is large, where probability 
 refers to the randomness in the source-destination set.
 Our main result is: for the $1 \over (1+d^\alpha) $ propagation model, 
 there exist constants $0<c, \overline{c}< \infty$ such that 
 for each $n$ there is a  $P_n > 0$ and for each source-destination 
 set $\sourcedest$, there is
 a system $\system(\sourcedest)$  with power $P_n$ and  throughput denoted 
 $\lambda_n(\sourcedest)$  
such that $\Pr \big(\system(\sourcedest) \mbox{ is successful}\big) \to 1$. 
In addition, if $\gamma < {1 \over 2}$,  
$\Pr \big(\lambda_n(\sourcedest)  \geq {W \over c n^{1-\gamma}} \big)  \to 1$, while
if $\gamma \geq {1 \over2}$, 
$\Pr \big( \lambda_n(\sourcedest)  \geq {W \over \overline{c} \sqrt{ n \ln n}} \big) \to 1$.
Note that the success and throughput of the system $\system(\sourcedest)$
depend on the locations of the source-destination pairs in $\sourcedest$.

\section{Propagation Model and Success Criterion Choices}
\label{sec-propagation}

In this section we indicate the specific propagation models and
 success criterion that we use in this paper.

\subsection{Propagation Models}

 In most of the prior work, e.g.
\cite{gupta:march2000}, the following
signal propagation model is adopted:

\vspace{.2em}
\begin{definition}  {\bf - Propagation Model A}:
\begin{eqnarray} \label{ch-propagation-eqn-model-A}
    \eta(d) = \frac{1}{d^\alpha},    \nonumber
\end{eqnarray}
where $\alpha > 0$ is a
constant whose value depends on the conditions of the channel.
\end{definition}
\vspace{.2em}
Notice that under Model A, when nodes become very close,
as happens for example when $n$ increases and $\gamma=0$ 
so that the network region $A_n$
remains fixed, the received power
will be larger than the transmitted, which is not reasonable.
In other words, Model A makes sense only as a far field assumption. 
This was noted by Arpacioglu and Haas in \cite{haas} and
by Dousse and Thiran in \cite{thiran:infocom04}.  In particular,
\cite{haas} considered the following alternative model:

\vspace{.2em}
\begin{definition}  {\bf - Propagation Model B}:
\begin{eqnarray} \label{ch-propagation-eqn-model-B}
    \eta(d)= \frac{1}{(1+d)^\alpha}.    \nonumber
\end{eqnarray}
\end{definition}
\vspace{.4em}
With this model, no
matter how close two nodes become, the received power is upper bounded
by the transmit power.  Similarly, \cite{thiran:infocom04} considered a
broad class of decreasing propagation models that are upper bounded
as $d$ approaches 0.

The difference between  models A and B 
has an important implication.  Consider a node $t$
transmitting to node $r$ and some other transmitting node $t'$
whose power received at $r$ appears as noise. Under Model A,
the ratio of powers received from each is
$\frac{\eta(\|t-r\|)}{\eta(\|t'-r\|)} =
\big( \frac{\|t'-r\|}{\|t-r\|}\big)^\alpha$.
As $\|t-r\|$ decreases, as long as $\|t'-r\|$ does not decrease as
fast, the ratio of $\eta(\|t-r\|)$ to $\eta(\|t'-r'\|)$ will increase.  
Thus if nodes only transmit to their closest neighbors,
interference from other transmissions (with similar transmit power)
will appear to be small by
comparison.  This potentially allows many simultaneous
transmissions throughout the network. On the other hand, 
under Model B
no matter how
close $t$ and $r$ become, interference from other transmissions can
be on a similar level. Therefore even if nodes transmit only to
their closest neighbors, interference from other transmissions may
still be significant.  This limits the number of
simultaneous transmissions, 
which in turn leads to different results on the throughput
scaling 
of a network.  For example, as mentioned in the introduction,
\cite{haas} showed that when the network region $A$ remains
fixed,  the per-node throughput under
Model B is  $\Theta({1 \over n})$, which is quite
different than the $\Omega \big({1 \over \sqrt{n \log n}} \big)$ 
found in \cite{gupta:march2000}.
This is precisely because the interference
prevents the number of simultaneous transmissions from growing to
infinity due to the boundedness of the received power under Model B,
so that nodes can only, in effect, use a time-division schedule.
A similar result was found in \cite{thiran:infocom04}
for all propagation models that are bounded at the origin.

On the other hand, if the underlying asymptotic regime is such that 
the node density is kept constant as $n$ increases, i.e. if $\gamma = {1 \over 2}$, 
then the difference 
between Model A and Model B discussed above will not effect the 
resulting scaling laws of the network (see Table \ref{tbl-summary}).

\subsection{Success Criterion}

In this paper,  we adopt the 
\emph{SINR (signal to interference and noise ratio)
criterion} \cite{gupta:march2000},  
which is commonly used for this purpose.  
To introduce it, consider the situation that a node at $t$ and all nodes at locations
in $T = \{t_1, \ldots, t_{M-1} \}$
transmit simultaneously in a given slot, that the transmission from $t$ is
intended to be received at $r$, that the received powers at $r$ from
the transmitters at $t$ and $T$ are $P'$ and $\{P_{1}, \ldots, P_{M-1}\}$, respectively,
and that background noise with power $N_o$ is also received.  
Let $\mathcal{T} = \mathbb{R}^2 \times \mathbb{R}^2 \times \overline{\mathbb{R}^2} $.

\begin{definition}{\bf - SINR} \\
The \emph{signal to interference noise ratio (SINR)} at $r$ is
\begin{equation}
	\mbox{SINR}\left( P', \{P_{1}, \ldots, P_{M-1}\}, N_o \right) 
	    = {  P' \over 
	      N_o + \sum_{i=1}^{M-1} P_{i}  }  ~.  \nonumber
 \end{equation}
When a propagation model $\eta$ is available, then with a small abuse of
notation, SINR becomes 
a function of  $(t,r,T) \in \mathcal{T}$ and $P$, as well as  $N_o$:
\[
         \mbox{SINR} \left( t, r, T, P, N_o, \eta \right) = \mbox{SINR} \left(
              P\eta(\|t-r\|), \{P\eta(\|t_1-r\|), \ldots, 
              P \eta(\|t_{M-1}-r\|) \}, N_o \right).
\]
\end{definition}

\vspace{.3em}
Note that with the above definition, transmissions at all locations in $T$ 
are considered noise.
We now use the above SINR functions to characterize 
a power indicator function
\[
\phi \left(P', \{P_{1}, \ldots, P_{M-1}\}, N_o \right) 
     =  \left\{ 
     \begin{array} {ll} 1, & \mbox{SINR}( P', \{P_{1}, \ldots, P_{M-1}\}, N_o) \geq \beta \\
               0, & \mbox{else}  \end{array} \right. ~,
\]
and the success indicator function based on $\phi$
\[
   \psi \left( t, r, T, P, N_o, \eta \right) = \left\{  \begin{array} {ll} 1, 
             & \mbox{SINR}( t, r, T, P, N_o, \eta) \geq \beta \\
               0, & \mbox{else}  \end{array} \right. ~.
\]
The success criterion to be used in this paper is that determined
by $\psi$ above, as summarized below. 
 
\vspace{.2em}
\begin{definition} {\bf - SINR Success Criterion}\\
Given $\beta>0$, $N_o > 0$,   $(t,r,T) \in \mathcal{T}$ and
propagation model $\eta$ (either Model A or B with associated
parameter $\alpha$),   
a transmission with power $P$ from $t$ to $r$  in the presence of
background noise with power $N_o$ and interfering 
transmitters at the locations in $T$, each with power $P$,  is said to be
\emph{SINR$_\beta$-successful at power $P$} (or we say $(t,r,T)$ 
\emph{satisfies the SINR$_\beta$ criterion at power $P$})  if 
\begin{eqnarray} \label{ch-propagation-tx}
   \mbox{SINR} \left( t, r, T, P, N_o, \eta \right) \geq \beta ,  \nonumber
\end{eqnarray}
This criterion is called the
{\em physical model} in \cite{gupta:march2000}.
\end{definition}

Note that in the next section, unless explicitly stated,
we do not restrict the 
transmitters and receivers to lie in any specified region such as a disk.

\section{Distance Based Success Criteria}
\label{sec-success}
To design a successful system $\system$, one must
design a set of compatible scheduled routes such that all induced hop sets 
$\hopset_1, \ldots, \hopset_p$ are successful with respect to the
SINR$_\beta$ criterion.  
While it is straightforward to check if any candidate hop set
is successful, it is not at all clear how one goes about designing a hop
set to be successful.  To facilitate such design, Gupta and Kumar
\cite{gupta:march2000} introduced a concept that we refer to as 
a \emph{distance-based} success criterion.  This is a criterion
that can be tested knowing only distances between transmitters, 
and between transmitters and receivers.
Specifically, Gupta and Kumar first introduced a
distance-based criterion   called the \emph{protocol model},
which is specified by two parameters 
$\rho$ and $\Delta$ and which  declares that a transmission from $t$
to $r$ is successful in the presence of other transmitters in $T$, i.e.,
$\psi(t,r,T,P,N_o,\eta)=1$,  
if $\|t-r\| \leq \rho$ and  $\|t'-r\| \geq \rho(1+\Delta)$ for all $t'
\in T$.  However, in deriving constructive results,
\cite{gupta:march2000} used  a distance-based criterion  of the
following form, which we find more useful.

\vspace{.2em}
\begin{definition} {\bf - Distance-Based Success Criterion   $\DC(C,D)$}\\
Given $C, D>0$, the transmission from $t$ to $r$ in the presence
of transmissions from the locations in the finite set  $T$ is said to be 
$\DC(C,D)$-successful 
(or we say $(t,r,T) \in \mathcal{T}$ satisfies the  $\DC(C,D)$ criterion)  if
\[ 
	\|t-r\| \leq C \nonumber 
\]
and
\[
	\|t'-t''\| \geq C(2+D)   \,  \mbox{ for all } t',t'' \in T
	\cup \{t\}.  \nonumber
\]
\end{definition}
\vspace{.3em}

Notice that there is no dependence on power, only on internode 
distances.  Notice also that instead of requiring  $\|t'-r\|$  to be large 
for $t' \neq t$  (as in the protocol model), this criterion requires  
$\|t'-t''\|$  to be large.  However, the triangle inequality implies 
that  if  $ t,r,T$  satisfies the  $\DC(C,D)$ criterion, then  $\|r-t'\| > C(1+D)$  
for all $t' \neq t$.  Moreover, the $\DC(C,D)$  criterion
has the effect of constraining the density of transmitters in $T$,
for reasons to be explained shortly.

Since the SINR$_\beta$ criterion is the preferred success criterion that 
we actually wish each system to satisfy, but a DC$(C,D)$ criterion is 
one that we can tractably design systems to satisfy, 
we will want to choose $C$  and $D$  so that 
the SINR$_\beta$ criterion is satisfied whenever the $\DC(C,D)$  
criterion is satisfied.  In this case,  $(C,D)$ are said to \emph{ensure}
the SINR$_\beta$ criterion, as defined below.

\begin{definition}
Given $N_o$ and $\eta$, a pair $(C,D)$  
is  said to \emph{ensure the SINR$_\beta$} criterion under propagation
model $\eta$, if there exists  $P>0$  such 
that any  $(t,r,T) \in \mathcal{T}$  that satisfies the $\DC(C,D)$ criterion
also satisfies the SINR$_\beta$ criterion at power  $P$, i.e  $\psi(t,r,T,P,N_o)=1$.
\end{definition}

The reason that $\DC(C,D)$  criterion constrains 
the distance between every pair of nodes in $T \cup \{t\}$, as opposed to
the distance between each transmitter in $T$ and $r$, is that it
limits the density of transmitters in $T$, which in the context of the
SINR$_\beta$ criterion, limits the total interfering power at $r$, and
enables a $\DC(C,D)$ criterion to ensure SINR$_\beta$.
  
The following lemma, whose proof uses techniques similar to those used in 
\cite{gupta:march2000}, will be used later to find  $(C,D)$
that ensure SINR$_\beta$.   From now on, unless otherwise stated,
we assume Propagation Model B, $\eta(d) = {1 \over (1+d)^\alpha}$.  
A similar result could be obtained for Propagation model A.

\vspace{.2em}
\begin{lemma}  \label{lem:SINRlowerbound}
For given  $N_o>0$, $\alpha>0$, and the
corresponding $\eta$ given by Propagation Model B, if $(t,r,T) \in \mathcal{T}$
satisfies the $\DC(C,D)$ criterion,  then for all $P>0$, 
\begin{equation} \label{eq:sinr_lower_bound}
	\mbox{SINR} \left(t,r,T,P,N_o,\eta \right)  \geq     {1 \over 
	      (1+C)^\alpha \left(  {N_o  \over  P}
	       +   \sum^{K}_{k=1} {6k+3 \over (1+kC(1+D/2))^\alpha} \right)   }  ~,
\end{equation}
where $K = \left\lfloor { \max \{ \|t'-r \| : t' \in T \} \over C(1 + D/2) }\right \rfloor$.
\end{lemma}
\vspace{.4em}

\noindent {\bf Proof:}
We are given that $(t,r,T) \in \mathcal{T}$ satisfies DC$(C,D)$ 
and that a transmitter at $t$ 
wishes to transmit to $r$ with power $P$ in the presence of noise power 
$N_o$ and simultaneous transmitters, each with power $P$, at the locations in $T$.   
We prove the lemma by upper bounding the number of transmitters
in $T$ in circular rings centered at $r$, and using the propagation law
to upper bound the power received from them.  Accordingly, let
$\delta = C (1+D/2)$, and let  $T_k$ denote 
the subset of transmitters in $T $ whose distance from $r$ is larger 
than $k \delta$ and no larger than $(k+1) \delta$, 
$k \in  \{1,2, \ldots, K \}$, with
$K = \big \lfloor { \max \{ \|t'-r \| : t' \in T \} \over \delta } \big  \rfloor$.
From the DC$(C,D)$ criterion and the triangle
inequality it follows easily that no transmitter in $T$ lies
within $\delta$ of $r$.  
Note also that $K$ has been chosen large enough that every transmitter
in $T$ is included in one of the $T_k$'s.   

\begin{figure} 
\centerline{\includegraphics[width=2.5in]{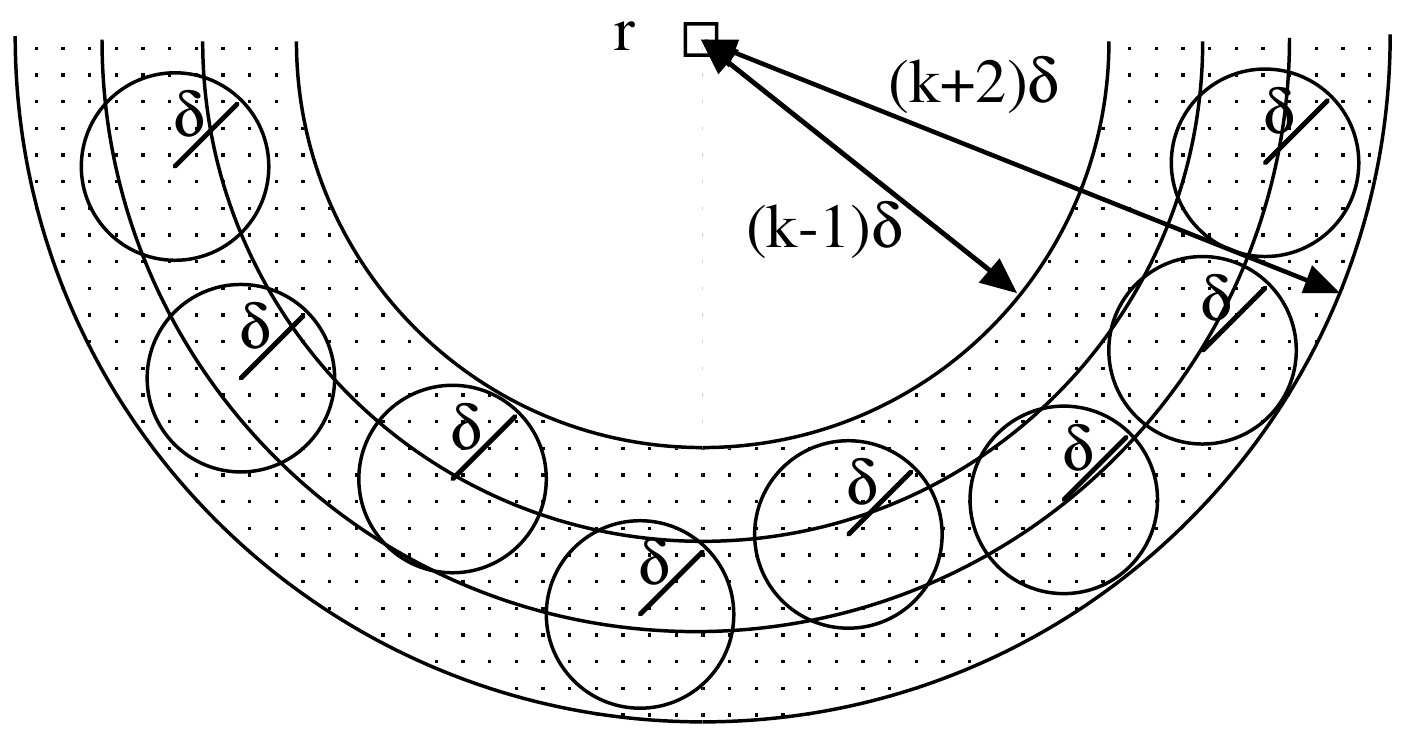}}
\caption{\label{fig:sinr-rings} The shaded region is the ring surrounding $r$
containing circles of radius $\delta$ centered on the transmitters in $T_k$.  }
\end{figure}

The number of transmitters in $T_k$, denoted $|T_k|$, 
can now be bounded from above by the area of the circular ring,
illustrated in Figure \ref{fig:sinr-rings}, with
outer radius $(k+2) \delta$ and inner radius $(k-1) \delta$ divided by 
the area of a circle of radius $\delta$.  This is because
the DC$(C,D)$ criterion implies that circles of radius $\delta$
centered about each transmitter in $T_k$ do not overlap, and because the 
circles corresponding to transmitters in $T_k$  lie within the aforementioned
ring.  It follows that
\begin{equation} \label{eq:Tk}
 	|T_k| \leq {\pi (k+2)^2 \delta^2 - \pi  (k-1)^2 \delta^2 
		\over \pi \delta^2 }   =  6k+3  ~. \nonumber 
\end{equation}
According to the propagation law, the power received at $r$ from a transmitter
in $T_k$ is at most  $P / (1 + k \delta)^\alpha$.  And since DC$(C,D)$
implies the power received from $t$ at $r$ is at least  $P/(1 +C)^\alpha$, we have
\begin{eqnarray}
	\mbox{SINR}(t,r,T,P,N_o,\eta) 
	   & = &  {  P\eta(\|t-r\|) \over 
	      N_o + \sum_{k=1}^{K} \sum_{t' \in T_k}  P\eta(\|t'-r\|) 	}
	  ~ \geq ~  { {P \over (1 +C)^\alpha }    \over
              N_o + \sum_{k=1}^{K} |T_k|  { P  \over  (1 + k \delta)^\alpha }   }  \nonumber \\
          & \geq & {1 \over
             (1+C)^\alpha \left( { N_o \over P} 
                   +  \sum_{k=1}^{K}  { 6k+3 \over  (1 + k \delta)^\alpha }  \right)  } ~,
                   \nonumber
\end{eqnarray}
which completes the proof of the lemma.    \hfill $\square$

\vspace{.5em}		
Notice that for any $C$ and $D$, one can choose $P$ so large that the term
$N_o / P$  in the denominator of ({\ref{eq:sinr_lower_bound}) is negligible.  
We therefore obtain the following.

\vspace{.2em}
\begin{corollary}  \label{lem:dc<beta}
$(C,D)$ ensures the SINR$_{\beta}$ criterion under Propagation Model B  if
\begin{equation}  \label{eq:criterion}
   (1+C)^\alpha \sum^{\infty}_{k=1} 
        {6k+3 \over (1+kC(1+D/2))^\alpha}  < {1 \over \beta} ~.
\end{equation}
\end{corollary}
\vspace{.5em}
This corollary can be strengthened somewhat if it is known 
that all transmitters and receivers lie in some bounded region $A$.  
Specifically, one can easily extend the proof to show that
for $(C, D)$ to ensure the SINR$_\beta$ criterion,
it suffices for (\ref{eq:criterion}) to hold with infinity as the upper
limit of the sum replaced by 
$\left \lfloor {  \mbox{\scriptsize diam}(A) \over C(1 + D/2) }\right \rfloor$.

The following lemma provides examples of $(C, D)$ that 
ensure the SINR$_\beta$ criterion.  

\vspace{.2em}
\begin{lemma} \label{lem:constantC}  
Let  $\beta >0$, $\alpha > 2$, and consider the corresponding Propagation Model B.  \\
\indent (a)  $(C,D)$ ensures SINR$_\beta$ if
\begin{equation}  \label{eq:1+CC}
	 { (1+C)  \over   C (2+D) }    < {1 \over \tau \beta^{1/\alpha} } 
\end{equation}
 \indent where $\tau = 2  \left( \sum^\infty_{k=1} {6 \over k^{\alpha-1} }   
         +  \sum^\infty_{k=1} {3 \over k^\alpha } \right)^{1 / \alpha}  $.\\
\indent  (b) For any $C > 0$,
there exists  $D >0$, depending on $\alpha$, $\beta$ and $C$ such that 
$(C,D)$ ensures\\
 \indent SINR$_{\beta}$.\\
\indent (c) There exists $D > 0$, depending on $\alpha$ and $\beta$,  
such that  $(C,D)$ ensures SINR$_{\beta}$
for all\\
\indent  sufficiently large $C$.
\end{lemma}

\vspace{.3em}
\noindent {\bf Proof:}  Let   $\alpha > 2$, $\beta > 0$.  
Dropping a ``1"  from the 
denominator in the left side of (\ref{eq:criterion})
yields the upper bound
\[ 
   (1+C)^\alpha \sum^\infty_{k=1}
                 {6k+3 \over (kC(1+D/2))^\alpha}    
       ~=~ { (1+C)^\alpha \over (C(2+D))^\alpha }  2^\alpha  \left(
                 \sum^\infty_{k=1} {6 \over k^{\alpha-1} }   
         +  \sum^\infty_{k=1} {3 \over k^\alpha } \right) ~.    
\]
Since $\alpha > 2$, both series on the right hand side are finite.  By
Corollary \ref{lem:dc<beta} if the above is less than $1 \over \beta$,
which is equivalent to (\ref{eq:1+CC}), then $(C,D)$ ensures SINR$_\beta$.
This shows (a).  
Part (b) follows directly from (a).  
Part (c) follows from (a) and the fact that for all sufficiently large 
$C$,  ${ 1+C  \over C } \leq  2  $.  
\hfill $\square$ 

\vspace{.5 em}
For large throughput, we would like $C$ and $D$ to be small
in order to permit a large number of  simultaneously transmitting nodes,
as will be evident in the proof of Theorem \ref{thm:main}.
On the other hand, they must ensure the SINR$_\beta$ 
criterion, and the next lemma demonstrates, not surprisingly, that when $C$ and $D$  are small, DC$(C,D)$ does not ensure the SINR$_\beta$ criterion.  Such
limitations on $C$ and $D$ are what limit the attainable throughput.
This lemma might also have future use in deriving upper bounds to
attainable throughput.  The proof is given in Appendix A.

\begin{lemma} \label{lem:converseDC}
Let $N_o$ and Propagation Model B with parameter $\alpha>2$ be given.  

\noindent (a) Given $C, D > 0$ and a positive integer $m$,
there exists  $(t, r, T)  \in \mathcal{T}$ 
such that $T$ has $m$ members,  DC$(C,D)$ is satisfied, and for any $P>0$,
\begin{equation} \label{eq:SINRupperbound}
     \mbox{SINR } \leq  \;  {1 \over 7}  \Big( {1 + 2C(2+D) \over 1+C } \Big) ^\alpha
         \Big( \sum_{k=1}^{\lfloor \sqrt{m/7} -2 \rfloor} {1 \over k^{\alpha-1}} \Big) ^{-1} ~.
\end{equation}
(b)  If  
\begin{equation}
     {1 \over 7}  \Big( {1 + 2C(2+D) \over 1+C } \Big) ^\alpha
         \Big( \sum_{k=1}^\infty  {1 \over k^{\alpha-1}} \Big) ^{-1}  ~<~ \beta
\end{equation}
then $(C,D)$ does not ensure SINR$_\beta$, i.e., there exists 
$(t,r,T) \in \mathcal{T}$  such that DC$(C,D)$ is satisifed but for all
$P >0$,  SINR$_\beta$ is not. \\
(c) If
 \begin{equation}
      {1 \over 7}   \Big( \sum_{k=1}^\infty  {1 \over k^{\alpha-1}} \Big) ^{-1} ~<~ \beta
\end{equation}
then all sufficiently small $C, D$ do not ensure SINR$_\beta$. 
\end{lemma}

\vspace{.5em}

In summary, the distance-based criterion DC$(C,D)$ can be used as
an intermediary that facilitates the design of systems that satisfy an
SINR$_\beta$ criterion.  To do so, one needs to choose $C, D$
appropriately, for example, as indicated in Lemma \ref{lem:constantC}.
The choice depends significantly on the propagation model.
For example, when $n$ nodes are distributed over a region $A$ with unit area,  
one may conclude from \cite{gupta:march2000} that $C_n = \sqrt{\ln n \over \pi n} $ 
and $D_n$ equal to an appropriate constant
ensures SINR$_\beta$  under Propagation Model A, whereas 
 \cite{duarte:2005} shows this does not hold under Propagation Model B.

\section{The Principal Result}
\label{sec-positive}

The following is the principal result of this paper.

\begin{theorem} \label{thm:main}
Consider the many-to-many communication task for a set
of $n$ source-destination pairs $\sourcedest$ randomly distributed
over a disk of 
radius $n^\gamma$, $\gamma \geq 0$, with a propagation model of
the form $\eta(d) = {1 \over (1+d)^\alpha}$ with $\alpha > 2$, an SINR$_\beta$ 
success criterion with parameter $\beta > 0$,  and background noise with power $N_o$.  
Then there exist constants $c_{\ref{thm:main}}, \overline{c}_{\ref{thm:main}} > 0$
depending only on $\alpha, \beta$ 
such that for any $n$ and any source-destination set $\sourcedest$, 
there exists a many-to-many system $\system(\sourcedest)$ and a power $P_n$ with 
throughput $\lambda_n(\sourcedest)$ such that for any packet transmission rate
$W>0$ (bits per slot),  as $n \rightarrow \infty$
\begin{equation} \label{eq:thmsuccess}
	\Pr \left( \system(\sourcedest) \mbox{  is SINR}_\beta \mbox{-successful} \right) 
	  \rightarrow 1    \nonumber
\end{equation}
and if $\gamma < {1 \over 2}$,
\begin{equation}   \label{eq:thmthroughputA}
	\Pr \Big( \lambda_n(\sourcedest) \geq {W  \over c_{\ref{thm:main}} n^{1-\gamma}} \Big) \rightarrow 1 ~,
\end{equation}
whereas if $ \gamma \geq {1 \over 2}$, 
\begin{equation} \label{eq:thmthroughputB}
	\Pr \Big( \lambda_n(\sourcedest) \geq {W  \over \overline{c}_{\ref{thm:main}} \sqrt{n \ln n} } \Big) \rightarrow 1 ~.
\end{equation}
When $\gamma < {1 \over 2}$,  $P_n$ depends on $\alpha,
\beta$, $N_o$, but not $n$, $\sourcedest$ or $\gamma$.
When $\gamma \geq {1 \over 2}$, $P_n$ depends on 
$\alpha$, $\beta$, $N_o$,  increases with $n, \gamma$, and 
does not depend on $\sourcedest$.     
\end{theorem}

\vspace{.5em}
To prove this theorem, we first prove a result like the above, but with
the SINR success criterion replaced by a distance-based 
success criterion.  The previous theorem will
then be proven by appropriate choices of the parameters
of the distance-based criterion. 

\begin{theorem} \label{thm:dist}
Consider the many-to-many communication task for a set
of $n$ source-destination pairs $\sourcedest$ randomly distributed
over a disk of radius $n^\gamma$, $\gamma \geq 0 $, with a 
distance-based success criterion.
For each $n$, let $C_n$ be chosen so that
\begin{equation} \label{eq:C<n}
		{C_n \over n^\gamma} <  { 1 \over 2} \; \mbox{ for all sufficiently large } n 
\end{equation}
and
\begin{equation}  \label{eq:C_over_n}
         a \, n \,      \Big( {C_n \over n^\gamma} \Big) ^2 
       +  \ln {C_n \over n^\gamma}  \rightarrow \infty 
\end{equation}
where $a = {1 \over 2^{13} \pi}  \ln {e \over 2}$.
Then for any $n$, any $D_n > 0$ and any source-destination set $\sourcedest$, 
there exists a many-to-many system $\system(\sourcedest)$, with throughput 
$\lambda_n(\sourcedest)$, such that for any packet transmission rate
$W>0$ (bits per slot), as $n \rightarrow \infty$
\begin{equation} \label{eq:thmsuccessdist}
   \Pr \left( \system(\sourcedest) \mbox{  is DC}(C_n,D_n) \mbox{-successful} \right)
	 \rightarrow 1 ~ 
\end{equation}
and 
\begin{equation} \label{eq:thmthroughputdist}
	\Pr \Big( \lambda_n(\sourcedest) \geq  {W \over c_{\ref{thm:dist}} \, n^{1-\gamma} 
	 \, C_n(2+D_n)^2} \Big) \rightarrow 1 ~.
\end{equation}
where $c_{\ref{thm:dist}} = 27 \times 2^{14} $.
\end{theorem}
\vspace{.5em}

Note that we have not attempted to minimize the constant $c_{\ref{thm:dist}}$.  Note also
that there is no restriction on $D_n$.  
Reducing $D_n$ permits  (\ref{eq:thmthroughputdist}) to guarantee a higher
throughput, up to a point of diminishing returns.  However,
when we apply this result to prove Theorem \ref{thm:main},
we will see, not surprisingly, that if $D_n$ is too small,
the distance-based success criterion will not ensure
the SINR$_\beta$ criterion. 

We now comment on conditions (\ref{eq:C<n}) and (\ref{eq:C_over_n}).  
The first places a natural upper bound on  $C_n$
as half the radius of the network region $A_n$.
For example, when $\gamma = 0$, it requires  
$C_n \leq {1 \over 2}$.
The second prevents $C_n$ from being too small. 
We note that the expression in (\ref{eq:C_over_n}) is a monotonically
increasing function of $C_n$.  It is satisfied, for example, 
if  ${C_n \over n^\gamma} = b \,  \big( {\ln n \over n} \big) ^{1/2} $ and
 $b \geq { 1 \over \sqrt{2a}}$, but not if $b <  {1 \over \sqrt{2a}}$.
When applying Theorem \ref{thm:dist} to prove
Theorem \ref{thm:main},  it turns out that (\ref{eq:C_over_n}) 
comes into play only when $\gamma \geq {1 \over 2}$.
In addition, the proof of Theorem \ref{thm:dist} will demonstrate 
that conditions (\ref{eq:C<n})
and (\ref{eq:C_over_n}) are sufficient to ensure that
as $n$ increases, with probability approaching one, the wireless network formed 
by the nodes is  \emph{connected}, in the sense that there is a route from every
node to every other node with all hops having length $C_n$ or less.
Gupta and Kumar \cite{gupta:k:98}  found a necessary and sufficient 
condition for such asymptotic connectivity that, when the network
region $A_n$ has radius $n^\gamma$, reduces to
$n \big( { C_n \over n^\gamma } \big)^2 - \, \ln n \rightarrow \infty$.  
Since (\ref{eq:C<n}) and (\ref{eq:C_over_n}) are sufficient to ensure 
asymptotic connectivity, they
must of course imply the complete connectivity condition  of \cite{gupta:k:98}, as well.  
We make a direct verification of this in Appendix B.  We also note that  
${C_n \over n^\gamma} = b \,  \big( {\ln n \over n} \big) ^{1/2} $
satisfies the connectivity condition for every $b>1$, whereas (\ref{eq:C_over_n})
is satisfied only if $b$  is at least ${ 1 \over \sqrt{2a}}$, which is much larger
than one.
However, this difference is not significant in determining the rate of
throughput scaling.  

In the remainder of this section we prove Theorem \ref{thm:dist};
then use the latter to to prove Theorem \ref{thm:main}.  We also describe
how it can be used to derive the $\Omega \big( {1 \over \sqrt{n \ln n} } \big)$
scaling result of \cite{gupta:march2000}.

\vspace{.5em}
\noindent {\bf Proof of Theorem \ref{thm:dist}:}
We follow an approach similar in many respects to that of \cite{gupta:march2000}.  
Given $ \gamma \geq 0, W, n$, a source-destination set  $\sourcedest$, 
and $C_n, D_n$ satisfying (\ref{eq:C<n}) and (\ref{eq:C_over_n}), we 
need to construct a system that with high probability is successful and has
the desired throughput.
The proof is divided into steps.  The first three describe a procedure
for designing a system for a particular $n$ and $\sourcedest$; the remaining
steps derive the performance of the designed system. 
Specifically,
Step 1 chooses a route for each source-destination pair, 
i.e.~a sequence of hops from the source to the destination.
This is done in a way that will make it possible to show that
all hops have length less than or equal to $C_n$,  with 
probability approaching one as $n$ tends to infinity, where probability is
with respect to the random source-destination set.
Moreover, these routes are chosen in a way that attempts to
limit $L_n$, the maximum number of routes assigned to any one node.
Step 2 chooses a collection of potential transmitter sets $T_1, \ldots, T_{S_n}$
such that the members of each set are at least $C_n(2+D_n)$ apart from 
one another (so that according to the DC$(C_n,D_n)$ criterion,
 they may simultaneously and successfully transmit to receivers 
 at distance $C_n$ or less), 
and every node is included in exactly one set (so that every node 
is permitted to transmit).
These sets are chosen with the goal of minimizing $S_n$.
Step 3 combines the routes of Step 1 and the potential transmitter sets
of Step 2 to form compatible scheduled routes, 
i.e.~a system, with period $p_n = L_n S_n $ and throughput
$\lambda_n  = {W \over L_n S_n}$. 
This system will be DC$(C_n,D_n)$-successful if and only if all hops have length 
$C_n$ or less, as was the goal of Step 1.
Step 4 shows this to be the case with probability approaching
one as $n \rightarrow \infty$.  In addition, it shows that 
$L_n = \Om( {n^\gamma \over C_n})$ with probability approaching one.
Step 5 shows that  $S_n = \Om \left( n^{1-2\gamma} C_n^2(2+D_n)^2 \right)$  
with probability approaching one as $n \rightarrow \infty$. 
Step 6 completes the proof by using the results of 
Steps 4 and 5 to show (\ref{eq:thmsuccessdist}) and (\ref{eq:thmthroughputdist}).  

\vspace{.5em}
{\bf \noindent Step 1: Route selection}

Given $n$ and $C_n$,  let $z = { C_n \over 2}$,
and partition the network region 
region $A_n$, which is a disk of radius  $n^\gamma$,  into 
convex cells, each having diameter at most $z$ and area 
at least $\mu z^2 $, where $\mu>0$ is some constant that does
not depend on $n$, $\gamma$, or $z$.  
While it is intuitively clear that this can be done, 
Lemma \ref{lem:partition} of Appendix C provides a concrete proof
with $\mu = {1 \over 512}$.  
It requires the radius   $n^\gamma$  to be at least  $z \over 4$, which holds
because of the choice $z = {C_n \over 2}$ and the assumption
(\ref{eq:C<n}) that $C_n < {n^\gamma  \over 2} $.
The number of cells in the partition, denoted $M_n$, is at most
${\pi n^{2\gamma} \over \mu z^2}$.

For each $(s,d)$ in the source-destination set $\sourcedest$,  
draw a straight line from $s$  to  $d$.  
Form a route for this pair by following the line
from $s$ to $d$ and selecting one node, and a corresponding hop, 
from each cell intersected by the line, whenever there is such a node.  
Convexity of the cells ensures that the line does not pass through the
same cell twice.   The fact that cells have diameters no 
larger than $z$ implies that if there is at least one node in each cell
intersected by the line, then the length of each hop is at most  
$2 z =  C_n$.  
Ordinarily, there will be more than one node in a given cell, a fact
that can be used to reduce the likelihood that a node is assigned to too 
many routes. Indeed, if $X_i$ source-destination lines intersect the $i$th cell 
and this cell contains $Y_i$ nodes, 
we apportion the $X_i$ routes as equally as possible among the $Y_i$ nodes.  
Thus, each node is assigned to no more than $\left\lceil {X_i \over Y_i} \right\rceil$ routes.
If there are no nodes in $i$th cell,  then we take $\left\lceil {X_i \over Y_i} \right\rceil$ 
to be infinity, even if $X_i = 0$.   
(We will show later that  $\Pr( \min_i Y_i = 0) \to 0$ as $n \to \infty$.)
Let $L_n$ denote the maximum of  $\left\lceil {X_i \over Y_i} \right\rceil$ over all cells.

\vspace{.5em}
{\bf \noindent Step 2:  Potential transmitter sets}

We begin by forming a graph with 
the $n$ nodes as the vertices and an edge between any pair 
of nodes separated by  $ C_n (2+D_n)$ or less. 
Let  $S_n$ denote the maximum number
of edges connected to any one node. 
We use the graph coloring theorem \cite{chartrand:1985,bondy:1976}
to assign one of $S_n$ distinct colors to each node in such a way
that no two nodes connected by an edge receive the same 
color\footnote{Actually, $S_n-1$ colors are sufficient, but  we use $S_n$
to simplify expressions.}.
We then partition the $n$ nodes into $S_n$ transmitter sets
$T_1, \ldots, T_{S_n}$ according to their assigned colors.
Since each node in one transmitter set is separated by 
$C_n (2+D_n)$ from every other node in the set, simultaneous transmissions by all
nodes in the set to receivers 
located within $C_n$ of each transmitter will be DC$(C_n,D_n)$-successful. 

\vspace{.5em}
{\bf \noindent Step 3:  System}

We now form a system $\system(\sourcedest)$ with period $p(n) = L_n S_n$. 
Recall the routes found in Step 1 and the groups of potential transmitters, 
$T_1, \ldots, T_{S_n}$, found in Step 2. 
Consider the sequence of $L_nS_n$ transmitters sets  
$\overline{T}_1, \ldots, \overline{T}_{L_n S_n} = T_1, \ldots, T_{S_n},
T_1, \ldots, T_{S_n}, \ldots, T_1, \ldots, T_{S_n}$.
We now schedule
the hops of the routes in ``rounds".  In the first round, for $j=1,\ldots, S_n$,
and for each node in $\overline{T}_j$, select a route 
to which it is assigned (if any) and schedule the corresponding hop from this 
node for time slot $j$.  In the second round, for 
$j=S_n+1,\ldots, 2S_n$, and for each node in $\overline{T}_j$,
select a route (if any) to which it is assigned that was not selected in 
the previous round, and schedule the corresponding hop from this note 
for time slot $j$.  
We continue in this way for $L_n-2$ further rounds.  Since each node
of each route is assigned to at most $L_n$ routes,
each of its assigned hops will be assigned a time slot.  
The resulting scheduled routes are compatible, because in any time slot each node
is assigned to only one route.  
In summary,  we have created a system $\system(\sourcedest)$ 
with period $p(n) = L_n S_n$ and throughput   
\[
	\lambda_n(\sourcedest) =  { W \over L_n S_n }~.
\] 
 
Note that the use of a partition and straight lines to define the routes
in Step 1 is just as in \cite{gupta:march2000}, except that we describe how 
to partition a disk, rather than the surface of a sphere.  
Unlike \cite{gupta:march2000},
we apportion the load as equally as possible among the nodes in each
cell. 
Because of  this, in Step 2 
we needed to color a graph with one vertex for each node, in contrast to
\cite{gupta:march2000}, which colored a graph with one vertex for each cell.
Since our graph has more nodes, it requires more colors, i.e., a larger
$S_n$.  However, the
analysis in Step 4 is simplified, because it is easier to determine which
nodes interfere with one another than which cells
interfere with one another.  Moreover, the throughput is not affected
because each node in our system 
is responsible for correspondingly fewer routes, i.e., a smaller $L_n$,  
than each cell in the system design of \cite{gupta:march2000}.

\vspace{.5em}
{\bf \noindent Step 4:  \boldmath $ L_n = \Om \big( {n^\gamma \over C_n} \big)$ 
with high probability }

Recall that $X_i$ denotes the number of source-destination lines  that intersect the
$i$th cell of the partition chosen in Step 1, and $Y_i$ denotes the number
of nodes in the $i$th cell. 
The following lemma provides a bound to $L_n = \max_i  \left\lceil { X_i  \over Y_i }\right\rceil$
that applies with probability approaching one.  Notice that when a source-destination
line intersects a cell, say $i$, that has no nodes, then $Y_i = 0$, and by
the convention of Step 1, $L_n=\infty$.  Therefore, the result of the lemma below
also implies that with probability approaching one, all hops of all routes are 
no longer than $2z = C_n$.  This, in turn, implies the connectivity mentioned
in the discussion after the theorem statement.  

\vspace{.5em}
\begin{lemma}   \label{lem:X/Y}   Under the conditions of Theorem \ref{thm:dist}, 
\[
	\Pr \Big( L_n \leq c_{\ref{lem:X/Y}} {n^\gamma \over C_n} + 1  \Big)  \rightarrow 1,
	~ \mbox{ as } n \rightarrow \infty
\]
where $c_{\ref{lem:X/Y}} = 3 \times 2^{13} \pi $. 
\end{lemma} 

\vspace{.5em}
\noindent {\bf Proof:}  
Since $L_n \defeq \max_i \left\lceil {X_i  \over Y_i} \right\rceil 
\leq {\max_i X_i \over \min_i Y_i} + 1$, 
we will separately consider the behavior of $\max_i X_i$ and $\min_i Y_i$,
finding quantities $a(n),b(n) > 0$ such  that $\max_i X_i$ exceeds
$a(n)$ with vanishing probability,
$\min_i Y_i$ is less than $b(n)$ with vanishing probability, 
and ${a(n) \over b(n)} \leq c_{\ref{lem:X/Y}} {n^\gamma \over C_n}$.

We begin by writing  $Y_i = \sum_{j=1}^n B_{i,j}$,  
where $B_{i,j} = 1$ when the $j$th node lies in the $i$th cell of the partition,
and  $B_{i,j} = 0$ otherwise.  
By the model for the random location
of nodes, for each $i$, $B_{i,1},\ldots, B_{i,n}$ are IID with 
\begin{equation}
    q_{n,i} \defeq \Pr(B_{i,j}=1) =  E[B_{i,j}]  = { 
{ \mbox{\scriptsize area of $i$th cell} \over \mbox{\scriptsize area of network region}}}
       \geq   {\mu C_n^2 \over 4\pi n^{2\gamma}} \defeq q_n ~.  \nonumber
\end{equation}
It follows that $E[Y_i]  \geq n q_n =   {\mu C_n^2 \over 4 \pi} n^{1-2\gamma}$.
Applying the union bound yields
\begin{equation}  \label{eq:unionY}
	\Pr \Big( \min_i Y_i <  {1 \over 2} n q_n \Big) 
             ~\leq~   \sum_{i=1}^{M_n} \Pr \Big( Y_i < {1 \over 2} nq_n \Big) 
\end{equation}

Similarly, write $X_i = \sum_{j=1}^n A_{i,j}$,  
where $A_{i,j} = 1$ when the line from $s_j$ to $d_j$ passes through the $i$th cell,
and  $A_{i,j} = 0$ otherwise.  By the model for the random location
of sources and random choices of destinations, for each $i$,
$A_{i,1},\ldots, A_{i,n}$ are independent and identically distributed (IID), with 
         $p_{n,i} \defeq  \Pr(A_{i,j}=1) = E[A_{i,j}]$.  
Note that the $X_i$'s are not identically distributed.
For example, $p_{n,i}$ is larger for a cell near the center of
the disk than one near the edge.  Nevertheless, Lemma \ref{lem:intersectprob}
of Appendix C finds a common upper bound to all $p_{n,i}$'s, namely,
\begin{equation}
      p_{n,i}  \leq 3  {C_n \over n^{\gamma}} \defeq  p_n  \nonumber
\end{equation}
for all $i$.  (Lemma \ref{lem:intersectprob} requires $z \leq n^\gamma$,
i.e.~$C_n \leq 2 n^\gamma$, which is guaranteed by (\ref{eq:C<n}).)
It follows that $E[X_i]  \leq n p_n =  3 C_n   n^{1-\gamma}$.
Once again we apply the union bound.
\begin{eqnarray}  \label{eq:unionX}
	\Pr \Big( \max_i X_i >  2 n p_n  \Big) 
		&\leq&   \sum_{i=1}^{M_n} \Pr \left( X_i > 2np_n \right) 
\end{eqnarray}

The facts that   $E[Y_i] \geq n q_n$ and  $E[X_i] \leq np_n$  and that
 $\min_i Y_i$ and $\max_i X_i$  have mean values 
in the vicinity of $n q_n$ and $n p_n$, respectively, 
suggests  that the probabilities appearing in the summations 
in  (\ref{eq:unionY}) and (\ref{eq:unionX})  are tail probabilities.
Accordingly, they can be effectively bounded above using Chernoff bound
techniques.  From, Lemma \ref{lem:chernoff} of Appendix C,
which uses the Chernoff bound, we have
\begin{equation}
      \Pr \Big(Y_i < {1 \over 2} n q_n \Big)    \leq  
       \exp \Big\{\!  -  n q_n \Big( {1 \over 2}  \ln {1 \over 2e}+1 \Big)  \Big\}
      ~=~ \exp \Big\{ \! - {1 \over 2} n q_n   \ln {e \over 2} \Big\}  \nonumber
\end{equation}
and
\begin{equation}  \label{eq:chernoffX}
	\Pr(X_i > 2n p_n) \leq  \exp \Big\{ \! -  n p_n \Big( 2 \ln {2 \over e}+1 \Big)  \Big\}
	~=~   \exp \Big\{\! -  n p_n  \ln {4 \over e}   \Big\}
\end{equation}
(Application of Lemma \ref{lem:chernoff} requires  $p_n <1$ and $q_n < 1$,
both of which are implied by (\ref{eq:C<n}).)

Substituting, the above into  (\ref{eq:unionY}) and  (\ref{eq:unionX}), 
respectively, gives
 \begin{eqnarray}  \label{eq:Ybound}
     \Pr \Big( \min_i Y_i <  {1 \over 2}n q_n \Big) 
           & \leq & \sum_{i=1}^{M_n}  \exp \left\{ - {1 \over 2} n q_n 
               \ln {e \over 2}  \right\} 
	   ~=~   \exp \Big\{\! - {1 \over 2} n q_n \ln {e \over 2} + \ln M_n   \Big\}    
	                    \nonumber \\
	 & \leq &   \exp \Big\{ \! -  {1 \over 2} n {\mu  C_n^2 \over 4 \pi n^{2\gamma}} 
	         \ln {e \over 2}
	           +   \ln  { 4 \pi n^{2\gamma} \over \mu  C_n^2}     \Big\}    \nonumber \\
	  &=&  \exp \Big\{ \! - 2 a n \Big( {C \over n^\gamma} \Big)^2 
	            -  2 \ln {C_n \over n^\gamma } +  \ln  { 4 \pi \over  \mu}     \Big\}    \nonumber \\
	  &\rightarrow& 0 \,,  \mbox{ as } n \rightarrow \infty ~.
\end{eqnarray}
where $a = {1 \over 2^{13} \pi } \ln {e \over 2}$,  and
\begin{eqnarray}   \label{eq:Xbound}
     \Pr \Big( \max_i X_i >  2 n p_n  \Big)  
           &\leq &     \sum_{i=1}^{M_n}  
               \exp \Big\{\! -n  p_n  \ln  {4 \over e }  \Big\} 
                     ~=~   \exp \Big\{ \! -n  p_n  \ln  {4 \over e } + \ln M_n  \Big\}
                      \nonumber \\
            &\leq&    \exp \Big\{ \! -n  {C_n \over n^{\gamma}}  3 \ln  {4 \over e } 
                + \ln  {4\pi n^{2\gamma}  \over \mu C_n^2}   \Big\}      \nonumber \\     
           &=&    \exp \Big\{ \!  -n  {C_n \over n^{\gamma}}  3 \ln  {4 \over e } 
            - 2  \ln {C_n \over n^\gamma} + \ln  {4 \pi  \over \mu }   \Big\}   \nonumber \\ 
            &\rightarrow & 0 \,,  \mbox{ as } n \rightarrow \infty     
\end{eqnarray}
where the convergence to zero in (\ref{eq:Ybound}) follows from condition
(\ref{eq:C_over_n}), and the convergence to zero in (\ref{eq:Xbound})
follows from (\ref{eq:C_over_n}) and the facts that  
$ {C_n \over n^\gamma} \geq  \big( {C_n \over n^\gamma}  \big) ^2$
(because (\ref{eq:C<n}) implies  ${C_n \over n^\gamma} < 1$) and
that $3 \ln {4 \over e} > 2 a$.

We now combine results.  With $c_{\ref{lem:X/Y}} = 3 \times 2^{13} \pi$, 
\begin{eqnarray}
    \Pr \Big( L_n \leq  c_{\ref{lem:X/Y}} {n^\gamma \over C_n} +1 \Big)
	 &=&   \Pr \Big( \max_i  \Big \lceil  {X_i \over Y_i} \Big \rceil
	   \leq  {2 n p_n  \over {1 \over 2} n q_n } +1 \Big)   \nonumber \\
	   &\geq&   \Pr \Big( \max_i {X_i \over Y_i} 
	   \leq {2 n p_n  \over {1 \over 2} n q_n } \Big)   \nonumber \\
	 &\geq&    \Pr \Big( \max_i X_i \leq 2 n p_n,  \, \min_i Y_i \geq 
	   {1 \over 2} n q_n  \Big)  \nonumber \\
	  & \rightarrow & 1 ~ \mbox{ as } n \rightarrow \infty ~,
\end{eqnarray}
where the convergence to one follows from  (\ref{eq:Ybound}) and 
 (\ref{eq:Xbound}).   This completes the proof of Lemma \ref{lem:X/Y}.   
\hfill $\square$

\vspace{1em}
{\bf \noindent Step 5:  \boldmath $S_n = 
\Om \Big(n^{1-2\gamma} C_n^2 (2+D_n)^2 \Big)$ with high probability}

Recall that  $S_n$  equals the largest number of edges to
which any node is connected.  By the definition of the graph, $S_n$
also equals the maximum, over all nodes, of the number of other nodes
within $C_n(2+D_n)$ of the given node. 
 
\vspace{.5em}
\begin{lemma}   
Under the conditions of Theorem \ref{thm:dist}, 
\begin{equation}
	\Pr \Big(  S_n \leq  {18 \over \pi } n 
	\Big( {C_n (2+D_n) \over n^\gamma } \Big)^2  \Big)
	~\rightarrow~ 1 ~ \mbox{ as } n \rightarrow \infty   ~. \nonumber
\end{equation}
\end{lemma}

\begin{figure}
\centerline{\includegraphics[width=1.5in]{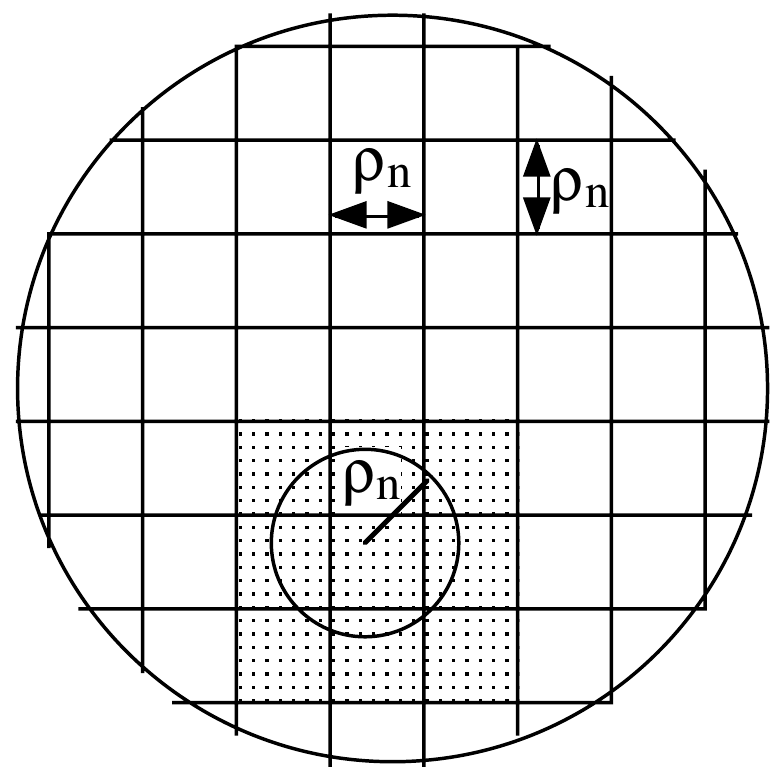}}
\caption{\label{fig:square-grid}  A circle of radius $\rho_n$
is contained in a shaded $3 \rho_n \times 3 \rho_n$ square.}
\end{figure}

\vspace{.5em}
\noindent {\bf Proof:}
Let $\rho_n \defeq C_n(2+D_n)$.
As illustrated in Figure \ref{fig:square-grid}, 
overlay a square grid with sides of length $\rho_n$ on the disk of 
radius $n^\gamma$.  This partitions the network region into cells
with area at most $\rho^2_n$.  The number of such cells, denoted
$\overline{M}_n$, is at most $  { \pi (n^\gamma+ \sqrt{2} \rho_n )^2  \over  \rho^2_n}  $, 
because the $\overline{M}_n$ squares of the grid that are contained in or
intersect the disk of radius $n^\gamma$ are all contained in a disk of
radius $n^\gamma + \sqrt{2} \rho_n $.  For $i=1,\ldots,\overline{M}_n$,
let   $U_i$  denote the number of nodes that lie in a $3\rho_n \times 3\rho_n$
square centered on the $i$th cell of the partition.   Since every circle
of radius $\rho_n$ lies in at least one of these  $3\rho_n \times 3\rho_n$
squares, it follows that  $S_n \leq \max_i U_i$.  Let us also observe
that $U_i = \sum_{j=1}^n B_{i,j}$, where $B_{i,j} = 1$ when the $j$th node
lies in the $3\rho_n \times 3\rho_n$ square centered on the $i$th cell 
of the partition, and $B_{i,j}=0$ otherwise.  By the model for the random
location of nodes, for each $i$, the $B_{i,j}$'s are IID with  
\begin{equation}
    p_{n,i} \defeq \Pr(B_{i,j}=1) =  E[B_{i,j}]  
    \leq { { \mbox{\scriptsize area of 9 squares centered on the $i$th cell} 
         \over \mbox{\scriptsize area of network region}}}
       =    {9 \rho^2_n \over \pi n^{2\gamma}} \defeq p_n ~.  \nonumber
\end{equation}
(Note that $B_{i,j}$, $p_{n,i}$ and $p_n$ are the not same as in the
proof of the previous lemma.)

Proceeding as in (\ref{eq:unionX}), (\ref{eq:chernoffX}) and (\ref{eq:Xbound}), we find
\begin{eqnarray}  \label{eq:PrSbound}
       \Pr \Big( S_n > {18 \over \pi} n^{1-2\gamma} \rho^2_n \Big)
          & < &   \Pr \Big( \max_i U_i  > 2 n p_n \Big)     
                     ~\leq~  \sum_{i=1}^{\overline{M}_n} \Pr( U_i > 2 n p_n) \nonumber \\
          & \leq &      \sum_{i=1}^{\overline{M}_n} \exp \Big\{\! -n p_n \ln {4 \over e} \Big\}
                    ~=~   \exp \Big\{ \! -n  p_n  \ln  {4 \over e } + \ln \overline{M}_n  \Big\}
                      \nonumber \\
            &\leq&    \exp  \Big\{\! -n  {9 \rho^2_n \over \pi n^{2\gamma}}  \ln  {4 \over e } 
                + \ln  { \pi (n^\gamma+ \sqrt{2} \rho_n )^2  \over  \rho^2_n}    \Big\}      
                	\nonumber \\     
          &=&    \exp   \Big\{ \! - 2 \Big( n \big( {\rho_n \over n^\gamma} \big) ^2  \,
                {9 \over 2 \pi}  \ln  {4 \over e } 
                \, - \,   \ln  \big( { n^\gamma \over \rho_n } + \sqrt{2} \big) \Big) + \ln \pi  \Big \}
                           \nonumber  \\   
            &\rightarrow & 0  ~ \mbox{ as } n \rightarrow \infty      
\end{eqnarray}
where the third inequality follows from Lemma \ref{lem:chernoff} and the fact that
$p_n \geq p_{n,i}$, and where the convergence is explained as follows.  
The principal terms in the last exponential above have the form 
$\overline{a} n v_n^2 -  \ln \Big( {1 \over v_n} + \sqrt{2} \Big)$, 
where $\overline{a} = {9 \over 2 \pi} \ln {4 \over e}$ and   $v_n = {\rho_n \over n^\gamma} = {C_n (2 + D_n) \over n^\gamma}
= u_n (2+D_n)$, and where $u_n = {C_n \over n^ \gamma} $.   
We now have
\begin{eqnarray} \label{eq:anvn_bound}
      \overline{a} n v_n^2 -  \ln \Big( {1 \over v_n} + \sqrt{2} \Big)  
          & \geq & \left\{  \begin{array} {ll} 
                          \overline{a} n v_n ^2 -  \ln {2 \over v_n} ~, & {1 \over v_n} \geq \sqrt{2} \\
                          \overline{a} n v_n ^2 -  \ln 2 \sqrt{2} ~, & \mbox{else}
                      \end{array}  
               \right.  \nonumber \\
          & = & \left\{  \begin{array} {ll} 
                          \overline{a} n v_n ^2 +  \ln  v_n -  \ln 2 ~, & {1 \over v_n} \geq \sqrt{2} \\
                          \overline{a} n v_n ^2 - \ln 2\sqrt{2} ~, &  \mbox{else}
                      \end{array}   
               \right. ~ \nonumber
\end{eqnarray} 
Since (\ref{eq:C_over_n}) shows $a n u_n^2 + \ln u_n \rightarrow \infty$, 
and since $v_n > u_n$, $\overline{a} > a$,  it follows that 
 $\overline{a} n v_n^2 +  \ln v_n \rightarrow \infty$.  Moreover,
 (\ref{eq:C_over_n}) implies $n u_n^2 \rightarrow \infty$; so $n v_n^2 \rightarrow \infty$,
 as well.  Using these facts in the above, shows that 
 $\overline{a} n v_n^2 -  \ln \Big( {1 \over v_n} + \sqrt{2} \Big) \to \infty$, which
establishes the convergence to 0 in (\ref{eq:PrSbound}), 
and completes the proof of the lemma.
\hfill $\square$

Note that in this step and the previous, we used the Chernoff bound to directly prove what was needed, rather than using the uniform convergence of the 
weak law of large numbers, as in \cite{gupta:march2000}.

\vspace{.5em}
{\bf \noindent Step 6: Completion of proof of Theorem \ref{thm:dist}}

The system $\system(\sourcedest)$ has been designed so that it will be successful provided only
that all hops have length $C_n$ or less, which, as explained just
before Lemma \ref{lem:X/Y}, 
happens if  $L_n \leq c_{\ref{lem:X/Y}} {n^\gamma \over C_n}$.  Therefore, from
Step 4,
\begin{equation}
    \Pr \left( \system(\sourcedest) \mbox{ is DC}(C_n,D_n)\mbox{-successful} \right)
    	~\geq~  \Pr \Big( L_n \leq c_{\ref{lem:X/Y}} {n^\gamma \over C_n} \Big)  ~\rightarrow~ 1 
	~\mbox{ as } n \rightarrow \infty ~.  \nonumber
\end{equation}
Since  $\lambda_n(\sourcedest) = {W \over L_n S_n}$, from Steps 4 and 5, and
the fact that $c_{\ref{thm:dist}} = c_{\ref{lem:X/Y}} {18 \over \pi}$, we have
\begin{eqnarray}
	 \lefteqn{ \Pr \Big( \lambda_n(\sourcedest) \geq  {W \over c_{\ref{thm:dist}} \, n^{1-\gamma} 
	 \, C_n(2+D_n)^2} \Big) ~~~~}     \nonumber \\
	 && ~~~~~= \Pr \left( L_n S_n \leq c_{\ref{thm:dist}} n^{1-\gamma} C_n (2+D_n)^2  
	     \right) \nonumber \\
       && ~~~~~\geq ~   \Pr \Big( L_n \leq  c_{\ref{lem:X/Y}} { n^\gamma \over C_n} 
         \mbox{ and }   S_n \leq {18 \over \pi}  n 
	\Big( { C_n (2+D_n) \over  n^{\gamma} } \Big)^2 \Big) \nonumber \\
     && ~~~~~ \rightarrow~ 1, ~ \mbox{ as } n \rightarrow \infty .  \nonumber
\end{eqnarray}
This completes the proof of Theorem \ref{thm:dist}.
\hfill $\square$

\subsection*{Proof of Theorem \ref{thm:main}}

Let $\alpha>2$, $\beta>0$ and $N_o >0$ be given.  
To prove this theorem, for each $n$ and for $\gamma$ in two
different ranges, we will make choices of $C_n, D_n$
so that the following hold:
(a)  $C_n$ satisfies (\ref{eq:C<n}) and (\ref{eq:C_over_n}), \, (b) 
$(C_n,D_n)$ ensures SINR$_\beta$ for all sufficiently large $n$,  
and (c)  $c_{\ref{thm:dist}} \, n^{1-\gamma} C_n (2+D_n)^2$
reduces (in the limit) to the expression in the denominator of 
(\ref{eq:thmthroughputA}) or (\ref{eq:thmthroughputB}), as
appropriate for the value of $\gamma$. 
Then for each $n$ and $\sourcedest$,  Theorem \ref{thm:dist}
will imply the existence of a system $\system(\sourcedest)$ that is 
DC$(C_n,D_n)$ successful with probability approaching one, 
whose throughput $\lambda_n(\sourcedest)$ satisfies
(\ref{eq:thmthroughputdist}).  
The fact that $(C_n, D_n)$ ensures SINR$_\beta$ for all
sufficiently large $n$ will imply 
the existence of a power $P_n$, depending on 
$\alpha, \beta, N_o, C_n, D_n, \gamma$, but not $\sourcedest$, 
such that the SINR$_\beta$ criterion is satisfied, whenever the DC$(C_n,D_n)$
criterion is satisfied.  Therefore, the fact that $\system(\sourcedest)$ is 
DC$(C_n,D_n)$ successful with probability approaching one will 
imply that $\system(\sourcedest)$ is SINR$_\beta$ successful 
with probability approaching one.  Finally, (\ref{eq:thmthroughputdist}) will imply
(\ref{eq:thmthroughputA}) or (\ref{eq:thmthroughputB}) as appropriate.

It remains to make choices for $C_n, D_n$.  
When $0 \leq \gamma < {1 \over 2}$, we choose $C_n = {1 \over 4}$,
which satisfies (\ref{eq:C<n}) and (\ref{eq:C_over_n}), 
and $D_n = \overline{C}_2$, where
$\overline{C}_2$ is a value, depending only on $\alpha$ and $\beta$,
such that DC$({1 \over 4},\overline{C}_2)$ ensures SINR$_\beta$, 
whose existence is established by Lemma \ref{lem:constantC} Part (b).
With $c_{\ref{thm:main}} = c_{\ref{thm:dist}} {1 \over 4} (2+\overline{C}_2)^2$, 
it follows immediately that $c_{\ref{thm:dist}} \, n^{1-\gamma} C_n (2+D_n)^2
= c_{\ref{thm:main}} \, n^{1-\gamma}$, so that (\ref{eq:thmthroughputA}) holds.  
Since $C_n, D_n$ are chosen to be constants, the power $P_n$ can be chosen
to be the same for all $n$ and all $\gamma \in [0, {1 \over 2})$.
Thus the proof of the theorem is complete for $\gamma < {1 \over 2}$. 

When $ \gamma \geq {1 \over 2}$, we choose 
$C_n = n^{\gamma - {1 \over 2}} \sqrt{ {2 \over a} \ln n }$,
$a = {1 \over 2^{13} \pi} \ln {e \over 2}$,
which satisfies (\ref{eq:C<n}) and (\ref{eq:C_over_n}).  
We also choose $D_n = \widetilde{D}$, where
$\widetilde{D}$ is a value, depending only on $\alpha$ and $\beta$, 
such that DC$(C,\widetilde{D})$ 
ensures SINR$_\beta$ for all sufficiently large $C$, 
whose existence is established by Lemma \ref{lem:constantC} Part (c).
With $\overline{c}_{\ref{thm:main}} 
= c_{\ref{thm:dist}} \sqrt{2 \over a} \, (2+ \widetilde{D})^2$, 
it follows immediately that $c_{\ref{thm:dist}} \, n^{1-\gamma} C_n (2+D_n)^2
= \overline{c}_{\ref{thm:main}} \, \sqrt{n \ln n}$, so that (\ref{eq:thmthroughputB}) holds.
When $\gamma \geq {1 \over 2}$, 
$C_n \to \infty$, and it follows that $P_n$ must also
tend to infinity as $n$ increases. Since  $C_n$ increases with
 $\gamma$, so too will $P_n$.   
This completes of the theorem for $\gamma \geq {1 \over 2}$.   
\hfill $\square$

\vspace{.5em}
We now comment on and justify the choices of
$C_n, D_n$ in the proof of Theorem \ref{thm:main}.  
Clearly, to maximize throughput we want to choose them
to minimize $C_n(2+D_n)^2$ while satisfying 
(\ref{eq:C<n}), (\ref{eq:C_over_n}), and the requirement
that $(C_n, D_n)$ ensure SINR$_\beta$.
Since we do not have a precise characterization of
the $(C_n,D_n)$ pairs that ensure SINR$_\beta$,
we simply try to make the most of 
the sufficient conditions in Lemma \ref{lem:constantC}.
Note that since we seek only to maximize the ``order" of the throughput,
we need not attempt a precise minimization.
To keep things simple, consider sequences $C_n$ that either 
decrease to zero, tend to a constant, or increase to infinity, and 
consider the same three possibilities for $D_n$.  

First, we cannot allow $C_n (2+D_n)$ to go to zero because
Lemma \ref{lem:converseDC} 
shows that in this case for large $n$,
$(C_n,D_n)$ will not ensure SINR$_\beta$.
Second, there is no point to making $D_n$ go
to zero, because the factor $(2+D_n)$ cannot decrease below 2.
Third, in the case of $\gamma < {1 \over 2}$,  
there is no point to having one of $C_n$, $D_n$ tend
to infinity while the other remains finite, 
because we can satisfy the constraints with both taking
finite values.   
It follows that for this case, the only potential competitor to the constant
$C_n, D_n$ that we chose in the proof of Theorem \ref{thm:main} is
$C_n \to 0$ and $D_n \to \infty$.
However, if $C_n \to 0$, then according to the sufficient
condition of  Lemma \ref{lem:constantC} Part (a),
we need  $D_n = \Omega \big( {1 \over C_n} \big)$, so that
$C_n (2+D_n)^2 = C_n \Omega \big( { 1 \over C_n^2 } \big) \to \infty$,
which of course is much worse than when $C_n, D_n$ are chosen
to be the constants in the proof of Theorem \ref{thm:main}.

Next in the case of $\gamma \geq {1 \over 2}$,
in addition to the first two points above, we note that one cannot satisfy 
(\ref{eq:C_over_n})
unless $C_n \to \infty$.  In the proof of Theorem \ref{thm:main} we chose
$C_n$ as small as possible and used Lemma \ref{lem:constantC}
Part (c) to justify a constant choice of $D_n$.  Making $D_n \to \infty$
would only reduce throughput.  Therefore, for both ranges of $\gamma$,
the choices of $C_n, D_n$
in the proof are as good as we can make them with the available
sufficient conditions for ensuring SINR$_\beta$. 

In summary, we note that among the constraints on $C_n, D_n$,
when $\gamma < {1 \over 2}$, it is the requirement for ensuring
SINR$_\beta$ that limits throughput, whereas when $\gamma \geq {1 \over 2}$,
it is the connectivity-ensuring requirement (\ref{eq:C_over_n})  that limits
throughput.

We conclude this section by noting that one can also use Theorem
\ref{thm:dist} to show straightforwardly that throughput 
$\Omega \big( {1 \over \sqrt{n \ln n} } \big)$ is attainable for propagation
model $1 \over d^\alpha$ and $\gamma \geq 0$.  This demonstrates
the original result of \cite{gupta:march2000}, as well as the fact that
it applies for all $\gamma \geq 0$.  
To do so, one lets $C_n = n^{\gamma - {1 \over 2}} \sqrt{ {2 \over a} \ln n }$,
$a = {1 \over 2^{13} \pi} \ln {e \over 2}$,
which satisfies (\ref{eq:C<n}) and (\ref{eq:C_over_n}), and one shows 
there is a $\widetilde{C}_2$ such that DC$(C_n,\widetilde{C}_2)$
ensures the SINR$_\beta$ criterion for the $1 \over d^\alpha$
propagation model.  Substituting  $C_n$ and $D_n = \widetilde{C}_2$ into
(\ref{eq:thmthroughputdist}) of Theorem \ref{thm:dist} yields throughput 
$\Omega \big( {1 \over \sqrt{n \ln n} } \big)$.

\section{Concluding Remarks}
\label{sec-conclusions}

In this paper we developed a constructive lower bound on attainable per-node
throughput in a wireless network whose nodes are randomly distributed over a disk, 
with radius  
growing as $n^\gamma$ with number of nodes $n$, for some $\gamma \geq 0$. 
By selecting $\gamma \in [0,\frac{1}{2})$, we can describe networks
ranging from fixed size to fixed density.  The lower bound has the form 
$\Omega \big( {1 \over n^{1-\gamma}} \big)$ when $\gamma < {1 \over 2}$, 
and $\Omega \big( {1 \over \sqrt{n \ln n} } \big)$ 
when $\gamma \geq {1 \over 2}$.
 
We now compare and contrast the approach
used to derive our results to those used by Gupta and Kumar in \cite{gupta:march2000}. 
First, recall that to prove Theorem \ref{thm:main},
we first proved Theorem \ref{thm:dist} for a distance-based success criterion
and then chose the constants $C_n$ and $D_n$ to permit
Theorem \ref{thm:dist} to guarantee the largest possible throughput, 
while ensuring the SINR criterion.
This is the strategy used in \cite{gupta:march2000}, except that 
it did not separate the derivation into two theorems, nor did it
separate the discussion of how a distance-based criterion (which
they called a protocol model) can ensure an SINR criterion, as we did
in Section \ref{sec-success}.  We view that separating 
into two theorems and separating the discussion of distance-based
criteria clarifies the derivation.  
We also indicated at the end of the previous section 
how the original $\Omega \big( {1 \over \sqrt{n \ln n} } \big)$
result of \cite{gupta:march2000} could be derived with our methods.

In \cite{gupta:march2000}, it was also shown that throughput of order 
$\Theta \big( {1 \over \sqrt{n \ln n} } \big)$
is the best attainable when $\gamma = 0$, $\eta(d) = {1 \over d^\alpha}$,
successful transmission occurs at rate $W$ or 0 depending on whether
received SINR is above or below a threshold $\beta$, and the system is
\emph{protocol based}, which in our terms means, essentially, that the system 
is designed so there are constants  $C$ and $D$ that
ensure the SINR$_\beta$ criterion such that all hops of
all routes have length $C$ or less, and all simultaneous transmitters
are at least $C(2+D)$ apart from each other.  Because of this
and because our system is protocol based and designed in a similar fashion,
it seems likely that the throughputs demonstrated by Theorem \ref{thm:main}
are also order optimal among protocol-based systems.  Indeed, for
$\gamma < {1 \over 2}$ they might be optimal among all systems,
because the system that attains this throughput is not limited by the connectivity 
condition (\ref{eq:C_over_n}) of Theorem \ref{thm:dist}, whereas the 
throughput $\Omega \big( {1 \over \sqrt{n \ln n} } \big)$ of \cite{gupta:march2000} 
is clearly limited by the analogous connectivity constraint. 
%




\section*{Appendix A}  

\setcounter{equation}{0}                        
\renewcommand{\theequation}{A\arabic{equation}} 

\setcounter{theorem}{0}                         
\renewcommand{\thetheorem}{A\arabic{theorem}}   


\noindent \textbf{Proof of  Lemma \ref{lem:converseDC}}:
(a)  Suppose we are given $\beta$, $N_o$, Propagation Model of B with parameter $\alpha>2$, $C,D > 0$ and a positive integer $m$.
Let $\delta = C (2+D)$.  Consider the following specific choice
of $(t,r,T)$.  Let  $r$ be at the origin, let $t$ be at distance $C$
from the origin, and as illustrated in Figure \ref{fig:smallC},
 for $k = 1,2,3, \ldots$, place as many nodes as possible 
on the circumference of a circle of radius $2 k \delta$,  
subject to the constraint that nodes are  
$\delta$ apart, except that the Euclidean distance between the first and last chosen
on the circle can be in $[\delta,2\delta)$.  Stop after placing a total of $m$ 
nodes into $T$.
Let $T_k$ denote the nodes on the circle with radius $2k\delta$, 
let $|T_k|$ denote the number of nodes in it.  
Let $K$ denote the number of rings into which we have placed nodes.
Notice that any two nodes, whether on the same circle or not, are at least
$\delta$ apart, except for $t$ and $r$, which are $C$ apart.  Therefore,
$(t,r,T)$ satisfies DC$(C,D)$.

\begin{figure}
\centerline{\includegraphics[width=2.5in]{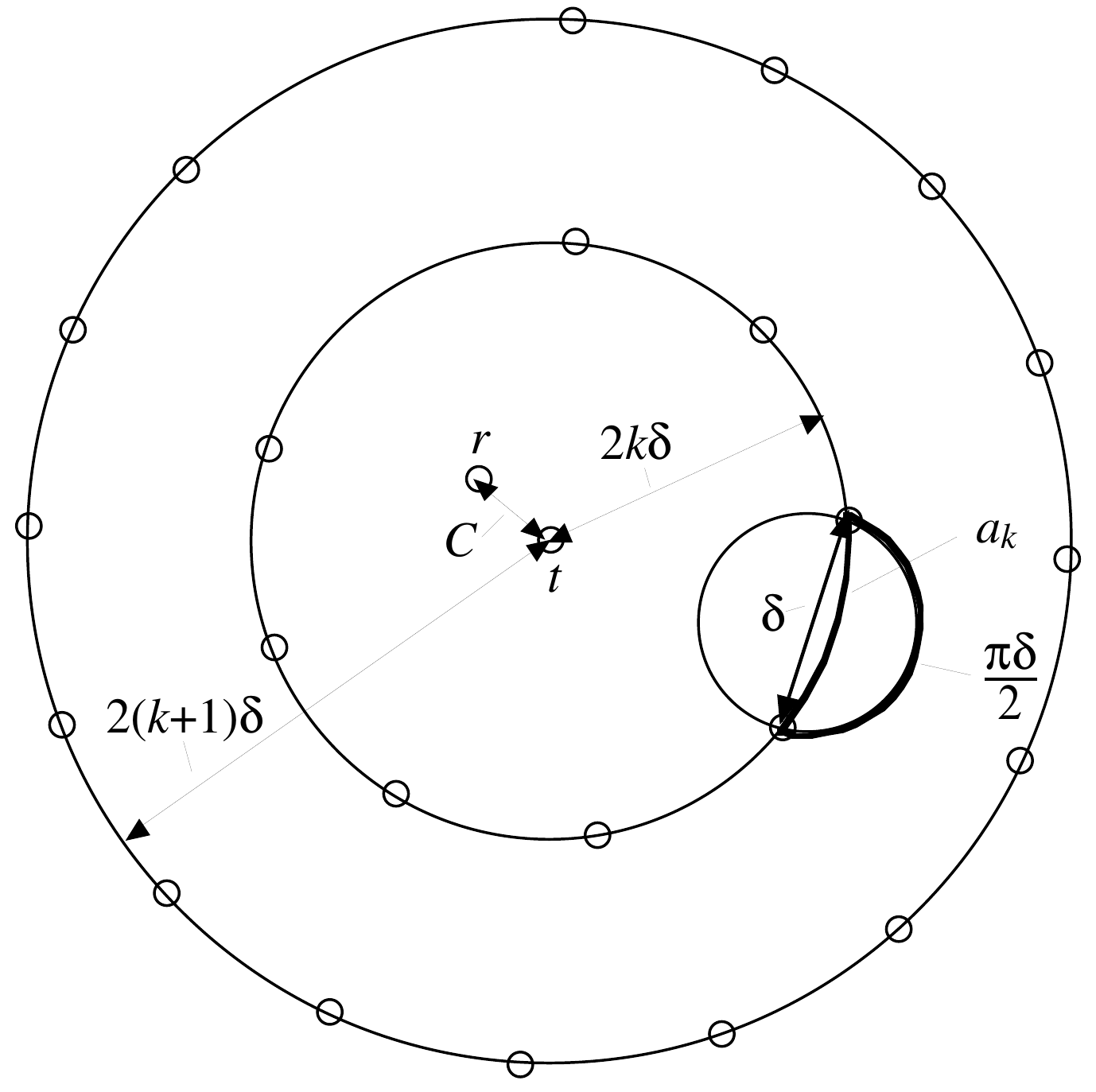}}
\caption{\label{fig:smallC}  Illustration of $t$, $r$
and the nodes in $T_k$ and $T_{k+1}$ for the proof of Lemma \ref{lem:converseDC}.}
\end{figure}

We now find bounds on $|T_k|$ and $K$.  Let $a_k$ denote the length
of an arc on the circumference of a circle of radius $2k \delta$ between two points 
that are $\delta$ apart.  Then
\begin{equation} 
         \delta ~ <  a_k ~ \leq ~ {1 \over 2} \pi \delta  \nonumber
\end{equation}
where the second inequality recognizes that the arc length
is bounded above by half the circumference of a circle with diameter $\delta$. 
Since     $|T_k| a_k \leq 2 \pi 2 k \delta$, it follows from the above that
\begin{equation}
           |T_k| ~\leq~  {4 \pi k \delta \over a_k }  ~<~ 4 \pi k  ~<~ 14 k  ~.    \nonumber
\end{equation}
Since for $k \leq K-1$,     $(|T_k|+1) a_k > 2 \pi 2 k \delta$, it follows that
\begin{equation}  \label{eq:Tlb}
       |T_k|  ~>~  {4 \pi k \delta \over a_k } - 1  ~\geq~  8 \pi k - 1  ~\geq~ 7 k  ~.
\end{equation}
To bound $K$, we observe that
\begin{equation}  \nonumber
     m  ~=~  \sum_{k=1}^K  |T_k|  ~<~  \sum_{k=1}^K  14k  
           ~=~  14 {K(K+1) \over 2}  ~<~  7 (K+1)^2 ~.
\end{equation}
Therefore, 
\begin{equation}  \label{eq:Klb}
          K ~ >~  \sqrt{m \over 7} - 1~.
\end{equation} 

We now bound SINR.  For any $P>0$,
\begin{eqnarray}
    \mbox{SINR}(t,r,T,P,N_o,\eta) 
	& = &  {  {P \over (1 + \| t-r \|) ^\alpha}   \over 
	  N_o + \sum_{k=1}^{K} \sum_{t' \in T_k}  {P \over (1 + \| t'-r \|) ^\alpha} }  
	               \nonumber \\
       &=&   { {P \over (1 +C)^\alpha }    \over   N_o +
              \sum_{k=1}^{K} |T_{k}|  { P  \over  (1 + 2k \delta)^\alpha }   }   
          ~ < ~  {1 \over (1+C)^\alpha} \,  { P    \over
               \sum_{k=1}^{K-1} 7 k  { P  \over  (1 + 2k \delta)^\alpha }   }   \nonumber \\
       & < &   {1 \over 7 (1+C)^\alpha} \,  { 1    \over
                 \sum_{k=1}^{K-1} { k  \over  (k + 2k \delta)^\alpha } }   
           ~ = ~   {1 \over 7} \Big( {1+2\delta \over 1+C} \Big)^\alpha \,  { 1    \over
                 \sum_{k=1}^{K-1} { 1 \over k^{\alpha-1} } }    \nonumber \\
      &<&   {1 \over 7} \Big( {1+2C(2+D) \over 1+C} \Big)^\alpha \,  { 1    \over
                 \sum_{k=1}^{\lfloor \sqrt{m \over 7} - 2  \rfloor} {1 \over k^{\alpha-1} } } 
                 \nonumber
\end{eqnarray}
where we used (\ref{eq:Tlb}) in the first inequality,  (\ref{eq:Klb}) in the third
inequality.
This proves Part (a).  Part (b) follows from Part (a) by choosing $m$ sufficiently large,
and Part (c) follows from Part (b) by choosing  $C,D$ sufficiently small.
\hfill $\square$



\section*{Appendix B}  

\setcounter{equation}{0}                        
\renewcommand{\theequation}{B\arabic{equation}} 

\setcounter{theorem}{0}                         
\renewcommand{\thetheorem}{B\arabic{theorem}}   

{\bf Condition (\ref{eq:C_over_n}) implies the connectivity condition of 
\cite{gupta:k:98}}

We show that condition (\ref{eq:C_over_n}), which requires
$a \, n \big( { C_n \over n^\gamma } \big)^2 + \, \ln {C_n \over n^\gamma}  
\rightarrow \infty$
implies the necessary and sufficient condition for connectivity, namely,  
\, $n \big( { C_n \over n^\gamma } \big)^2 - \, \ln n \rightarrow \infty$.
For brevity, let  $\rho_n = {C_n \over n^\gamma}$.  
We begin by using proof by contradiction to show that if  (\ref{eq:C_over_n}) holds, i.e. 
$a \, n \rho_n^2  + \ln \rho_n  \rightarrow \infty$, and $1 < b < \sqrt{1 \over a}$, 
then there exists $n_o$ such that $\rho_n > b \sqrt{ \ln n \over n}$ 
for all $n \geq n_o$.  (Recall that $a = {1 \over 2^{13} \pi } \ln {e \over 2} << 1$.)
Accordingly, suppose to the contrary that
for all $n_o$ there exists $n \geq n_o$ such that $\rho_n \leq b \sqrt{ \ln n \over n}$.
Let us choose $n_o$ large enough that  $b \sqrt{ \ln n_o \over n_o} < 1$.
Since $ { \rho^2 \over | \ln \rho | }  =   { \rho^2 \over - \ln \rho }$  is monotonic increasing for $\rho < 1$,  there exists $n \geq n_o$ such that
\[
a \, n \rho_n^2  + \ln \rho_n    ~\leq~  a \, b^2 \ln n + \ln b \sqrt{ \ln n \over n}  
         ~=~  \big( a b^2 - 1 \big) \ln n + {1 \over 2} \ln \ln n + \ln b ~. 
\]
Since (1) there are infinitely many $n$ for which the above holds, (2)
$(ab^2 -1) < 0$, and (3)  for large $n$ the right hand side above
above approaches $-\infty$, it follows
that  $\liminf_{n \to \infty} a \, n \rho_n^2  + \ln \rho_n = - \infty$, which  
contradicts the assumption that $n \rho_n^2  + \ln \rho_n  \rightarrow \infty$.  
Therefore,  there exists
$n_o$ such that $\rho_n >  b \sqrt{ \ln n \over n}$ for all $n \geq n_o$.
Since for all such $n$,  
$ n \rho_n^2 - \ln n > b \ln n - \ln n = (b-1) \ln n$,  and since $b > 1$,
we conclude that $n \rho_n^2 - \ln n \to \infty$, which is the desired result.



\section*{Appendix C}  

\setcounter{equation}{0}                        
\renewcommand{\theequation}{C\arabic{equation}} 

\setcounter{theorem}{0}                         
\renewcommand{\thetheorem}{C\arabic{theorem}}   

\begin{lemma}  \label{lem:partition}
For any $w>0$ and $\rho \geq 2w$,  there exists a partition of a disk of radius $\rho$ 
into convex cells such that each cell has diameter no larger than $8w$ and area at least $w^2/8$.
\end{lemma} 

Note that we have not attempted to make the bounds in this lemma as tight as possible.

\vspace{.5em}
\noindent {\bf Proof:}
Given $w>0$ and $\rho \geq 2w$, we will specify a set of points $G$ 
in the disk of radius $\rho$ and show that the Voronoi partition
corresponding to this set has the desired properties.
We will begin by choosing  $u >0$ and a positive integer $m$ such that 
$(m+1/2) u = \rho$  and $w \leq u < 2w$.  
Specifically,  choose $m$ such that $\rho = (m+1/2) w + r $, for some $r$,
$0\leq r < w$,  and choose $u = \rho/(m+1/2)$.  Since $\rho > 2w$,
it must be that $m \geq 1$.  Since 
$(m+1/2)w+r = (m+1/2) u$, we have  $w \leq u = w + r/(m+1/2) < w + w = 2w $.

To specify, $G$, we first place the center of the disk into $G$,
which we consider to be the origin of the Cartesian plane.
Next, for $d=1,\ldots,m$, add points on a ring of radius $d u$  to $G$
in such a way that the Euclidean distances from each point to its two immediate
neighbors on the ring are at least $u/2$ and no more than $u$. 
For $d=1$, one can simply add to $G$ the vertices of a regular hexagon
with sides of length $u$ centered at the origin.
For $d \geq 2$, start by placing one point, denoted $p_1$, arbitrarily on the 
ring.  Place a second point $p_2$ on the ring at distance $u/2$ from the first.  
To place the third point $p_3$, move on the ring away from $p_1$ and $p_2$ 
to a point at distance $u/2$ from $p_2$.   
Continue in this way to add points on the ring until the $n$th point $p_n$ is within 
distance $u/2$ of $p_1$.
Discarding $p_n$, one obtains the set of points $\{ p_1, \ldots, p_{n-1} \}$.  
Clearly every point in this set is at distance $u/2$ from both of its immediate neighbors,
except possibly for $p_1$ and $p_{n-1}$.  However,
$\| p_{n-1} - p_1 \| > u/2$, since $p_{n-1}$ was not the last
point picked, and by the triangle inequality 
$\| p_{n-1} - p_1 \| \leq  \| p_{n-1} - p_n \| + \| p_n - p_1 \|  \leq {u \over 2} + {u \over 2} = u$.  Thus, the set $\{ p_1, \ldots, p_{n-1} \}$
has the desired property that  the distances from each point to its immediate
neighbors on the ring are at least $u/2$ and no more than $u$.

Since points on distinct rings are at least $u$ apart, we see
that every point is at least $u/2$ apart from every other point.
Let $\Pi$ denote a Voronoi partition for this set.  That is, $\Pi$ is a partition 
with one cell for each point in $G$, and with each $x$ in the disk being 
contained in a cell corresponding to a point in $G$ to which it is closest.  
The cells of a Voronoi partition are convex.

Consider the Voronoi cell corresponding to some point $p \in G$.  For any $x$
in this cell, the distance to $p$ is at most $2u$, because the distance from $x$
to the closest spot (not necessarily a point in $G$)  on the ring containing $p$ 
is at most $u$, and 
because the distance from this spot to $p$ is at most $u$.  
Therefore, the diameter of the cell is at most $4u < 8w$.  
The fact that any two points are at least $u/2$ from each other
implies that every Voronoi cell contains a circle of radius $u/4$.  
Therefore, its area is at least $\pi u^2/ 16 > u^2 / 8 \geq  w^2 /8$.  
\hfill $\square$

\vspace{.5em}
\begin{lemma}  \label{lem:intersectprob}
For the partition chosen in Step 1, whose
cells have diameter $z$ or less, with $z \leq n^\gamma$,
\[ 
      p_{n,i}  \leq 6  {z \over n^{\gamma}}   
\]
for all $i$, where
 $p_{n,i}$  is the probability that the $i$th cell of the partition is 
intersected by a random line whose endpoints are independently drawn 
from the network region with uniform distributions.
\end{lemma}

\vspace{.5em}
\noindent {\bf Proof:}
We upper bound $p_{n,i}$ by the probability, denoted $\overline{p}_{n,i}$, 
of the cell being intersected by a random line after translating the cell
so as to maximize this probability.  The translated cell will contain 
the center of the circular network region, which we consider to be the
origin of the coordinate axes.
In turn, we bound $\overline{p}_{n,i}$ by bounding 
$\Pr(\mbox{line intersect}|r,\theta)$, 
which is the conditional probability of 
a line intersecting the translated cell given that $s$, the source end of the line,
is at $(r,\theta)$ in polar coordinates.  
For $r \leq 2z$, we use $\Pr(\mbox{line intersect}|r,\theta) \leq 1$.
For $r > 2z$, in which case $s$ cannot lie in the translated cell,
$\Pr(\mbox{line intersect}|r,\theta)$
equals the probability that the destination end of the random line $d$ lies in the 
shaded region shown.  The latter
is  bounded above by the area of the crosshatched  triangle shown 
in that figure, divided by the area of the network region.  Using the fact
that the cell diameter is at most $z$ we find
\[
       \Pr(\mbox{line intersect}|r,\theta) 
             ~\leq~ \left\{ 
                  \begin{array}{ll} 1, & r \leq 2z \\
                     { z {r + n^\gamma \over \sqrt{r^2 -z^2} } (r + n^\gamma)  \over  \pi n^{2\gamma} },
                     & r > 2z                  
                 \end{array}    
                  \right.
             ~<~  \left\{ 
                  \begin{array}{ll} 1, & r \leq 2z \\
                     { 2 z (r + n^\gamma)^2  \over  \sqrt{3} \pi r n^{2\gamma} },
                     & r > 2z                  
                 \end{array} 
                  \right.  ~.
 \]
Since the probability density of $(r,\theta)$  is  $p(r,\theta) = {2r \over n^{2\gamma}} {1 \over 2\pi}$,   we have
 \begin{eqnarray}
        p_{n,i} ~\leq~ \overline{p}_{n,i}  &=& \int_0^{2\pi}  \int_0^{n^\gamma}  
                  \Pr(\mbox{line intersect}|r,\theta) \,
		{2r \over n^{2\gamma}} {1 \over 2\pi} \, dr \, d\theta     \nonumber \\
	  &<&  \int_0^{2\pi}  \int_0^{2z} 1 \,  
		{2r \over n^{2\gamma}} {1 \over 2\pi} \, dr \, d\theta    
		   + \int_0^{2\pi}  \int_{2z} ^{n^\gamma}  
		   { 2 z (r + n^\gamma)^2  \over  \sqrt{3} \pi r n^{2\gamma} }
		{2r \over n^{2\gamma}} {1 \over 2\pi} \, dr \, d\theta   \nonumber \\
	&<&   4 { z^2  \over  n^{2\gamma} } +  {32 \over 3 \sqrt{3} \pi} {z \over n^{\gamma}}
	       ~<~ 6{z \over n^\gamma}   \nonumber
\end{eqnarray}
where the next to last inequality uses the fact that  $ {z  \over n^\gamma} < 1$.
\hfill $\square$

\begin{figure}
\centerline{\includegraphics[width=3.5in]{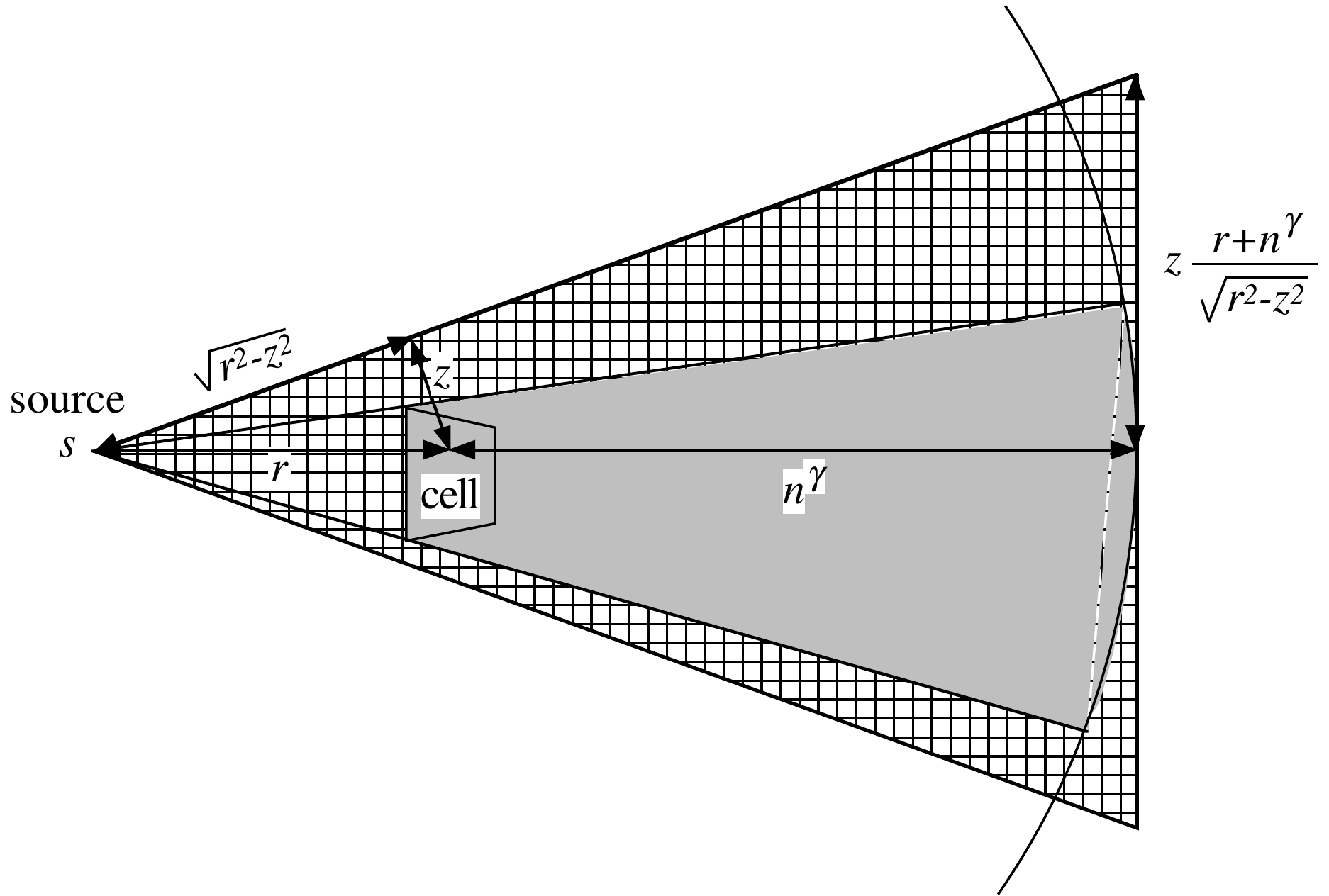}}
\caption{\label{fig:prob_intersect}  The route from a source $s$
to its destination $d$ passes through the displayed cell
if and only if $d$ lies in the shaded region.  The probability of this is 
bounded above by the probability that $d$ lies in the cross-hatched triangle.
The cell includes the center of the network region, which has been  
rotated relative  so that $s$ lies horizontally to the left of the origin.}
\end{figure}

\vspace{.5em}
\begin{lemma} \label{lem:chernoff}
Let $Y = \sum_{i=1}^n B_i$  be the sum of $n$  independent
and identical (IID) binary random variables 
$B_1, \ldots, B_n$, with $\Pr(B_i=1) = q = 1 - \Pr(B_i=0)$ and $0 < q < 1$.
Then for any $1 \leq \nu<1/q$
\begin{equation} \label{eq:Y>nunq}
	\Pr(Y > \nu n q) \leq  \exp \left\{ -  n q \left( \nu \ln {\nu \over e} + 1 \right) \right\}
\end{equation}
and for any $0 < \nu\leq1$
\begin{equation} \label{eq:Y<nunq}
      \Pr \left(Y < \nu n q \right) \leq  
                   \exp \left\{  -  n q \left( \nu \ln {\nu \over e} + 1 \right) \right\} ~.
\end{equation}
\end{lemma}

\vspace{.5em}
\noindent {\bf Proof:}  
Suppose $0 < q < 1$ and $1 \leq \nu < 1/q$.  Using the Chernoff bound
and the IID nature of the $B_i$'s, we have
\begin{eqnarray} \label{eq:Y>nunq2}
    \Pr(Y > \nu n q) &\leq& \min_{s \geq 0} \; E \left[ e^{s(Y-\nu n q)}  \right]
            ~=~  \min_{s \geq 0} \; \prod_{i=1}^n E \left[  e^{s(B_i-\nu q)} \right]  
             \nonumber   \\ 
             &=&  \Big(  \min_{s \geq 0} E \left[  e^{s(B_1-\nu q)} \right]  \Big)^n 
             ~=~   e^{-nD(\nu q || q) ) }
\end{eqnarray}
where $D(\nu q || q)$  denotes the divergence of the probability distribution
$\{\nu q, 1-\nu q\}$ with respect to the distribution $\{q,1-q\}$, which
is defined and bounded below:
\begin{eqnarray}  \label{eq:divergenceY}
     D(\nu q || q) 
	    &\defeq& \nu q \ln {\nu q \over q }
	                       + (1-\nu q) \ln {1-\nu q \over 1-q }
	   ~=~  \nu  q  \ln  {\nu q \over q }
	                 + (1-q) {1-\nu q \over 1-q } \ln {1-\nu q \over 1-q }  	
	                     \nonumber \\
	  &\geq& \nu  q  \ln  {\nu q \over q }
	             +  (1-q) \Big( {1-\nu q \over 1-q } -1 \Big)   
	  ~=~ \nu  q  \ln \nu   +  q  -\nu  q  
	           ~=~  q \left( \nu \ln {\nu \over e} + 1 \right)   \nonumber
\end{eqnarray}
where the inequality in the above uses  $x \ln x \geq x-1$.
Substituting the above into (\ref{eq:Y>nunq2}) gives (\ref{eq:Y>nunq}).

Now suppose $0 < q < 1$ and $0 < \nu \leq 1$.  Using the Chernoff bound
and the  IID nature of the $B_i$'s, we have
\begin{eqnarray}
    \Pr(Y < \nu n q) &\leq& \min_{s \geq 0} \; E \left[ e^{-s(Y-\nu n q)}  \right]
          ~=~  \min_{s \geq 0} \; \prod_{i=1}^n E \left[  e^{-s(B_i-\nu q)} \right] 
          	\nonumber \\
           &=&  \Big(  \min_{s \geq 0} E \left[  e^{-s(B_1-\nu q)} \right]  \Big)^n 
             ~=~   e^{-nD(\nu q || q) ) } ~.
\end{eqnarray}
From here, the proof is the same as for the previous case.
\hfill $\square$

\bibliographystyle{IEEE}
\bibliography{bibpropagation}

\end{document}